\documentclass[fleqn,usenatbib]{mnras}

\usepackage{newtxtext,newtxmath}

\usepackage[T1]{fontenc}

\DeclareRobustCommand{\VAN}[3]{#2}
\let\VANthebibliography\thebibliography
\def\thebibliography{\DeclareRobustCommand{\VAN}[3]{##3}\VANthebibliography}

\usepackage{graphicx}	
\usepackage{caption}
\usepackage{subfigure}
\usepackage{xspace}

\newcommand{\co}{$^{12}\mathrm{CO}$ \xspace}
\newcommand{\col}{$^{13}\mathrm{CO}$ \xspace}
\newcommand{\eddy}{\textsc{eddy} \xspace}
\newcommand{\discminer}{\textsc{discminer} \xspace}

\title[Dynamical mass measurements of protoplanetary discs]{Dynamical mass measurements of two protoplanetary discs}

\author[G. Lodato et al.]{
G. Lodato,$^{1}$\thanks{E-mail: giuseppe.lodato@unimi.it}
L. Rampinelli,$^{1}$
E. Viscardi,$^{1}$
C. Longarini,$^{1}$
A. Izquierdo,$^{2}$
T. Paneque-Carre\~{n}o,$^{2,3}$
\newauthor
L. Testi,$^{2,4}$
S. Facchini,$^{1,2}$
A. Miotello,$^{2}$
B. Veronesi$^{5}$
and C. Hall,$^{6}$
\\
$^{1}$Dipartimento di Fisica, Universit\`a degli Studi di Milano, Via Celoria 16, 20133 Milano, Italy\\
$^{2}$European Southern Observatory, Karl-Schwarzschild-Strasse 2, D-85748 Garching bei München, Germany\\
$^{3}$ Leiden Observatory, Leiden University, PO Box 9513, 2300, RA, Leiden, The Netherlands\\
$^{4}$INAF - Osservatorio Astrofisico di Arcetri, Largo E. Fermi 5, I-50125 Firenze, Italy\\
$^{5}$Univ Lyon, Univ Lyon1, Ens de Lyon, CNRS, Centre de Recherche Astrophysique de Lyon UMR5574, F-69230, Saint-Genis,-Laval, France\\
$^{6}$Center for Simulational Physics, The University of Georgia, Athens, GA 30602, USA
}

\date{Accepted XXX. Received YYY; in original form ZZZ}

\pubyear{2022}

\begin{document}
\label{firstpage}
\pagerange{\pageref{firstpage}--\pageref{lastpage}}
\maketitle

\begin{abstract}
ALMA observations of line emission from planet forming discs have demonstrated to be an excellent tool to probe the internal disc kinematics, often revealing subtle effects related to important dynamical processes occurring in them, such as turbulence, or the presence of planets, that can be inferred from pressure bumps perturbing the gas motion, or from detection of the planetary wake. In particular, we have recently shown for the case of the massive disc in Elias 2-27 how one can use such kind of observations to measure deviations from Keplerianity induced by the disc self-gravity, thus constraining the total disc mass with good accuracy and independently on mass conversion factors between the tracer used and the total mass. Here, we refine our methodology and extend it to two additional sources, GM Aur and IM Lup, for which archival line observations are available for both the $^{12}$CO and the $^{13}$CO line. For IM Lup, we are able to obtain a consistent disc mass of $M_{\rm disc}=0.1 \, \mathrm{M}_{\odot}$, implying a disc-star mass ratio of 0.1 (consistent with the observed spiral structure in the continuum emission) and a gas/dust ratio of $\sim 65$ (consistent with standard assumptions), { with a systematic uncertainty by a factor $\simeq2$ due to the different methods to extract the rotation curve.} For GM Aur, the two lines we use provide slightly inconsistent rotation curves, that cannot be attributed only to a difference in the height of the emitting layer, nor to a vertical temperature stratification. Our best fit disc mass measurement is $M_{\rm disc}=0.26 \, \mathrm{M}_{\odot}$, implying a disc-star mass ratio of $\sim 0.35$ and a gas/dust ratio of $\sim 130$. Given the complex kinematics in the outer disc of GM Aur and its interaction with the infalling cloud, the CO lines might not well trace the rotation curve and our results for this source should then be considered with some caution. 
\end{abstract}




\section{Introduction}

Accurate measurements of the total disc mass is one of the `holy grails' in protostellar disc studies. A knowledge of this fundamental quantity is essential for many different aspects in star and planet formation: (i) it determines how much mass is available for the formation of planets \citep{Manara18,Testi22}; (ii) it allows a comparison of disc evolutionary models to observations \citep{Lodato17,Tabone22,Manara22}; (iii) it determines the level of coupling between gas and dust, thus heavily influencing radial dust drift and dust trapping into pressure maxima \citep{Veronesi19,Powell19}; (iv) if the disc is massive enough, the disc self-gravity might lead to the development of gravitational instabilities, which will appear as spiral structures and lead to efficient angular momentum transport \citep{KratterLodato}. 

Yet, such a fundamental quantity is really hard to measure. The main constituent of the disc, the H$_2$ molecule, does not possess a permanent electric dipole and consequently it does not emit significantly. We are thus forced to use appropriate tracers to infer the disc mass (see \citealt{Miotello22} for a review). Traditionally, disc masses have often been derived from measurements of the \emph{dust} mass, following the pioneering studies of, e.g., \citet{Beckwith90}, by measuring the mm-continuum flux and translating it into a dust mass \citep{Ansdell16,Barenfeld16,Testi22}, by assuming an educated guess for the flux to mass conversion factor. This method has significant uncertainties though, as the conversion factor encapsulates all uncertainties related to the optical depth (e.g. see \citealt{Hartmann06}), on the dust opacity assumptions and on the gas/dust ratio (further details on the uncertainties in dust mass measurements can be found in \citealt{Manara22}). Alternatively, one might use the CO molecule, in its optically thin isotopologues $^{13}$CO and C$^{18}$O, as a tracer of the gas mass. Also this tracer, however, leads to very uncertain measurements. Indeed, CO-derived disc masses generally appear to be much lower than dust-derived masses \citep{Miotello16,Miotello17}, hinting at significant carbon depletion. Another clue that dust flux might be more directly related to the total disc mass rather than CO flux also comes from the fact that dust masses appear to correlate better with stellar masses and especially with mass accretion rates with respect to CO masses \citep{Manara16}. Finally, in a handful of sources the gas mass has been estimated from HD flux (an isotopologue of the most abudant H$_2$ molecule). Interestingly, in the case of TW Hya \citep{Bergin13}, the HD derived mass is consistent with the mass inferred from dust flux measurements with standard assumptions, while being significantly higher than that obtained from CO measurements, highlighting the difficulties of inferring gas masses from CO.

In this context, finding a probe to the disc mass that is independent of the specific tracer used would thus represent a great advance in our understanding of the structure, dynamics and evolution of planet forming discs. In \citet{Veronesi21} we have demonstrated a proof-of-concept method that is based on a precise determination of the rotation curve of the disc, from which we can obtain \emph{a dynamical measurement} of the disc mass, by fitting the observed curve to models including the disc contribution to the gravitational potential. In \citet{Veronesi21} we have applied this method to a peculiar source, Elias 2-27, that on the one hand provides an ideal case to test for self-gravity effects, given its estimated high dust mass, but on the other hand also poses some problems in the determination of an accurate rotation curve, due to foreground contamination of the CO lines used to obtain the rotation curve and due to its complex morphological structure \citep{Paneque21}. Yet, in that case we were able to measure dynamically the disc mass, obtaining a mass of $\sim 0.08\, \mathrm{M}_{\odot}$ and a gas/dust ratio very close to the standard value of 100 \citep{Veronesi21}. In this paper, we apply the same technique to two additional sources, IM Lup and GM Aur, for which exquisite kinematical data are available from the MAPS survey \citep{Oberg21}. We also refine our method in several ways: (i) we retrieve the rotation curve using different techniques, thus testing the robustness of the curve and of the methods used to obtain it; (ii) we include in a much more consistent way the contribution of the pressure gradient in the disc, as a function of the height of the relevant gas emitting layer; (iii) we explore the dependence of our results on additional structural parameters, such as the disc pressure scale-height. 

The paper is organized as follows. In section \ref{sec:sample} we describe the two sources we use for our analysis and the relevant data sets and parameters. In Section \ref{sec:methods} we describe our method to extract and model the rotation curve. In Section \ref{sec:results} we present and discuss our results on the disc mass of IM Lup and GM Aur. In Section \ref{sec:conclusions} we draw our conclusions. 

\section{Sample and data}

\label{sec:sample}

In order to test our method to fit for the disc mass, based on the rotation curve, we have chosen two sources for which there is tentative evidence of a relatively massive disc, IM Lup and GM Aur for which high quality kinematical data are available from the MAPS survey \citep{Oberg21}. 

\subsection{IM Lup}

IM Lup has been the subject of intense study over the years. It is a K5 star, located at a distance of 158 pc \citep{Gaia18}, with a stellar mass that was estimated to be $\approx 1\, \mathrm{M}_{\sun}$ \citep{panic09} (rescaled due to updated distance) based on SMA $^{12}$CO emission. It hosts an unusually large disc, extending out to $\approx 300$ au in the dust continuum and out to $\approx 1000$ au in the gas \citep{Cleeves16}. The dust emission shows a clear evidence of an extended spiral morphology, that may be indicative of gravitational instability, given the lack of evidence for either internal (planetary) or external companions \citep{Huang18}. 

Gas kinematics has been analysed by \citet{panic09}, based on SMA data (with a $1.8\times1.2$'' beam size), revealing an approximately Keplerian disc around a $\approx 1\, \mathrm{M}_{\sun}$ star. \citet{Cleeves16} estimate the disc properties by comparing SED and observed mm visibilities to a simple model of a tapered power-law density profile. They provide a crude estimate for the tapering radius of $R_{\rm c}=100$ au. The total dust mass is $1.7~10^{-3} \, \mathrm{M}_{\sun}$, which would imply a high gas mass of $0.17\, \mathrm{M}_{\sun}$, using a gas/dust ratio of 100. \citet{Pinte18} analyse their CO ALMA line emission data (beam size $\approx 0.5$'') with MCFOST obtaining a gas/dust ratio of 347, which would imply an excessively high gas mass of $0.6\, \mathrm{M}_{\sun}$, for standard CO abundances. Their model also indicates a tapered power-law density structure with a tapering radius $R_{\rm c}=284$ au. More interestingly, they provide a detailed analysis of the disc rotation: while the inner disc is in good agreement with Keplerian rotation around a ($1\pm 0.1\mathrm{)M}_{\sun}$ star, both the $^{12}$CO and the $^{13}$CO rotation curve appear to become significantly sub-Keplerian beyond $\approx 300$ au, which \citet{Pinte18} attribute to the effect of the steepening pressure gradient outside the tapering radius of the density profile.

In this paper we will use the $^{12}$CO (2-1) and $^{13}$CO (2-1) observations of IM Lup, recently observed within the ALMA Large Program MAPS \citep{Oberg21}, with a spatial resolution of $\approx 0.15$'' and a spectral resolution of 0.1 km/sec { and 0.2 km/sec respectively, and velocity sampling of 0.2 km/sec \citep{Oberg21} (Note that, as mentioned by \citealt{Oberg21} ``For the delivered data products the velocity resolutions have been coarsened to achieve more uniformity between spectral-line cubes as described in \citealt{Czekala21}'').} \citet{Zhang21} provide an analysis of the disc structure somewhat similar to that provided by \citet{Cleeves16}: they fit the SED, mm-band ALMA images and the CO emission surface to their MAPS data and obtain a best fitting value of 100 au for the tapering radius $R_{\rm c}$ and a total gas mass even higher than estimated by \citet{Cleeves16}, obtaining $M_{\rm d}\approx 0.2\, \mathrm{M}_{\sun}$.  

Such a high inferred disc mass should clearly show up in high quality kinematical data as a super-Keplerian contribution to the rotation curve, making this source a prime case for testing such effect.

\subsection{GM Aur}

GM Aur is another well-known disc, at a distance of 159 pc \citep{Gaia18}. The stellar mass has been estimated dynamically to be $\approx 0.84\, \mathrm{M}_{\sun}$ by \citet{Simon2000} based on IRAM low resolution $^{13}$CO emission, later corrected to $\approx 1.1\, \mathrm{M}_{\sun}$ by \citet{Dutrey08}, { when using the correct distance to the source} again from  kinematics \citep{Guilloteau14}, as well as spectral type fitting using pre-main-sequence tracks \citep{Macias18}. GM Aur is a transition disc, with an inner cavity of $\approx 35$ au and an extended dust disc, reaching out to $\approx 300$ au, and displaying evidence for substructure \citep{Macias18}. The dust mass estimated by \citet{Macias18} is $0.002\, \mathrm{M}_{\sun}$, which would translate into $0.2\, \mathrm{M}_{\sun}$ for the gas mass, assuming a gas/dust ratio of 100. The same value for the total gas mass is also obtained by \citet{Schwarz21}, based on thermo-chemical modeling of their MAPS line emission data. \citet{Schwarz21} also obtain a value for the tapering radius of $R_{\rm c}\approx 111$ au. 

MAPS data, { with a spatial resolution of $\approx 0.15$'' and a spectral resolution of 0.1 km/sec and 0.2 km/sec for $^{12}$ and $^{13}$CO respectively, and velocity sampling of 0.2 km/sec},  also show evidence of interaction of the GM Aur disc with the environment. Outwards of a roughly Keplerian disc out to $\approx 550$ au, \citep{Huang21} reveal a complex structure including a spiral arm and extending out to $\approx 1900$ au. Such extended structure might indicate late infall from the envelope and could justify the high mass inferred for this system. Naturally, these structures are visible also in moment one map, showing a wiggle-like structure in the observed velocity field, that could make the retrieval of the rotation curve more difficult (see below) \citep{Hall20,Longarini21,Terry22}.


\section{Methodology}

\label{sec:methods}

\subsection{Basic system parameters}

The general problem of fitting kinematical data to a model involves a large number of parameters, including geometrical parameters (e.g. distance, inclination, position angle of the source...), physical parameters (e.g temperature and pressure scale height, disc and stellar mass, etc.) and radiative parameters (location of the surface from where the different lines emit), depending on the sophistication of the model (see \citealt{Pinte22} for a recent review). Rather than constructing a single fitting procedure to fit for all of the unknown parameters (which would typically imply a large degree of degeneracy if the dataset is limited), here we  assume some of the system parameters from the literature (e.g., geometry, height of the emitting layers,...) and  fit for the stellar and disc masses based on an observed rotation curve. We will also fit for the disc size, since the effect of an exponential truncation in the disc does affect both the disc gravitational field and the pressure gradient contribution to the rotation curve (see also \citealt{Dullemond20}). 

Distances and inclinations of the two sources are obtained from the literature. In particular, we adopt the same geometrical parameters as used in MAPS \citep{Oberg21}, we fix the pressure scale height from the thermochemical model of \citet{Zhang21} { (which are in very good agreement with the recent analysis of \citealt{Paneque22})} and the height of the emitting layers of $^{12}$CO and $^{13}$CO from the MAPS analysis \citep{Law21}. 

In general, we assume that the disc midplane temperature profile is described as a single power-law with index $-q$:
\begin{equation}
    T(R)=T_0\left(\frac{R}{R_0}\right)^{-q},
\end{equation}
where $R_0$ and $T_0$ are a scale radius and the corresponding temperature, respectively. The pressure scale height in hydrostatic balance is thus given by
\begin{equation}
    H(R)=\frac{c_{\rm s}}{\Omega_{\rm K}}=H_0\left(\frac{R}{R_0}\right)^{n},
\label{eq:thickness}
\end{equation}
where $c_{\rm s}\propto T^{1/2}$ is the sound speed, $\Omega_{\rm K}=(GM_\star/R^3)^{1/2}$ is the Keplerian velocity at radius $R$ and $n=(3-q)/2$. $H_0$ is the value of the scale height at $R_0$ and only depends on the temperature $T_0$. 

For the line emitting scale height, we assume that it is given by a tapered power law for each CO isotopologue:
\begin{equation}
    z_i(R)=z_{0,i}\left(\frac{R}{R_0}\right)^{p_i}\exp\left[-\left(\frac{R}{R_{{\rm t},i}}\right)^{q_i}\right],
\end{equation}
where the index $i$ refers to the two isotopologues used here ($^{12}$CO and $^{13}$CO), $z_{0,i}$ is a scaling factor, $R_{{\rm t},i}$ is the tapering radius and the power law indices $p_i$ and $q_i$ define the shape of the emitting layer. As reference radius, we choose $R_0=100$ au. 
Table \ref{tab:basicparameters} shows the values that we assume for all the parameters mentioned in this section.

\begin{table}
 \caption{Basic parameters of the systems considered in this work.}
 \label{tab:basicparameters}
 \begin{tabular}{l c c}
  \hline
  & IM Lup & GM Aur \\
  \hline
  $d$ (pc) & 158 & 159 \\
  $i$ (degrees) & 47.5$^\circ$ & 53.2$^\circ$ \\
  \hline
  $H_0$ (au) & 10 & 7.5 \\[2pt] 
  $q$ & 0.66 & 0.3 \\[2pt]
  $n$ & 1.17 & 1.35 \\[2pt]
  \hline
  $z_{\rm 0,12CO}$ (au) & 163.9 &  37.34\\[2pt]
  $p_{\rm 12CO}$ & 3.144 & 1.066 \\[2pt]
  $q_{\rm 12CO}$ & 0.655 & 4.988 \\[2pt]
  $R_{\rm t,12CO}$ (au) & 40.13 & 598.95 \\[2pt]
  \hline
  $z_{\rm 0,13CO}$ (au) & 7.65 & 2.19 \\[2pt]
  $p_{\rm 13CO}$ & 2.599 & 4.539 \\[2pt]
  $q_{\rm 13CO}$ & 4.993 & 4.98\\[2pt]
  $R_{\rm t,13CO}$ (au) & 304.6 & 237.9\\[2pt]
  \hline
 \end{tabular}
\end{table}

\subsection{Disc radial and vertical structure}

We assume that the disc surface density is described as a tapered power law, following the similarity solutions of \citet{Lyndenbellpringle}:
\begin{equation}
    \Sigma(R)=\frac{(2-\gamma)M_{\rm d}}{2\pi R_{\rm c}^2}\left(\frac{R}{R_{\rm c}}\right)^{-\gamma}\exp\left[-\left(\frac{R}{R_{\rm c}}\right)^{2-\gamma}\right],
    \label{eq:sigma}
\end{equation}
where $R_{\rm c}$ is the disc size and $M_{\rm d}$ is the disc mass, that we will set as free parameters to be fitted for. In the following, we will fix $\gamma=1$.

For what concerns the vertical disc structure, we assume that the disc is vertically isothermal and that the density and pressure profiles are given by:
\begin{equation}
\rho(R,z)=\rho_0(R)\exp\left[-\frac{R^2}{H^2}\left(1-\frac{1}{\sqrt{1+z^2/R^2}}\right)\right],
\label{eq:density}
\end{equation}
\begin{equation}
P(R,z)=P_0(R)\exp\left[-\frac{R^2}{H^2}\left(1-\frac{1}{\sqrt{1+z^2/R^2}}\right)\right],
\label{eq:pressure1}
\end{equation}
where $\rho_0$ and $P_0$ are the midplane density and pressure, respectively, and the thickness $H$ is given by Eq. (\ref{eq:thickness}). Note that if the disc is locally isothermal then $P/\rho=P_0/\rho_0=c_{\rm s}^2$.
For $z\ll R$ the above expressions reduce to the standard Gaussian profiles often used to describe the structure of a vertically isothermal disc.

One might wonder if assuming a Gaussian profile for the vertical structure is appropriate for the relatively massive discs that we want to model here. It is well known that for a self-gravitating disc the vertical structure is described by an hyperbolic cosecant function, with thickness { given by \citep{KratterLodato}}
\begin{equation}
    H_{\rm sg}=\frac{c_{\rm s}^2}{\pi G \Sigma}.
\end{equation}
This is correct only in the limit where the disc dominates the gravitational potential, a situation that does not occur in our case. The vertical structure starts to be affected by self-gravity effects when the usual axisymmetric stability parameter $Q=c_{\rm s}\kappa/\pi G \Sigma\approx 1$ (where $\kappa$ is the epicyclic frequency). This may or may not be the case for our sample (although we anticipate here that the effects of self-gravity on the rotation curve start becoming significant much before the disc reaches marginal stability, $Q\approx 1$, see below). In the general case where both the star and the disc contribute to the gravitational field there is no analytical exact solution. However, it can be shown \citep{BL99,KratterLodato} that the vertical structure is still approximately described by a Gaussian, with a thickness that departs from eq. (\ref{eq:thickness}) only by factors of order unity (depending on the disc surface density). This provides only a higher order correction to the rotation curve and we thus neglect it here.

\subsection{Extracting the rotation curve}

To test the robustness of our model we obtained the rotation curves of the two sources using both \eddy \citep{teague2019eddy} and \discminer \citep{Izquierdo21}, each of which has got its advantages and drawbacks. We can then fit the rotation curves for our parameters and analyse how the systematic errors of the rotation curve retrieval procedure affect the results. For a direct comparison of rotation curves extracted with the methods presented below, we use the disc orientation and emission surface parameters summarised in Table 1.

\begin{figure*}
    \centering
    
     \includegraphics[width=\columnwidth]{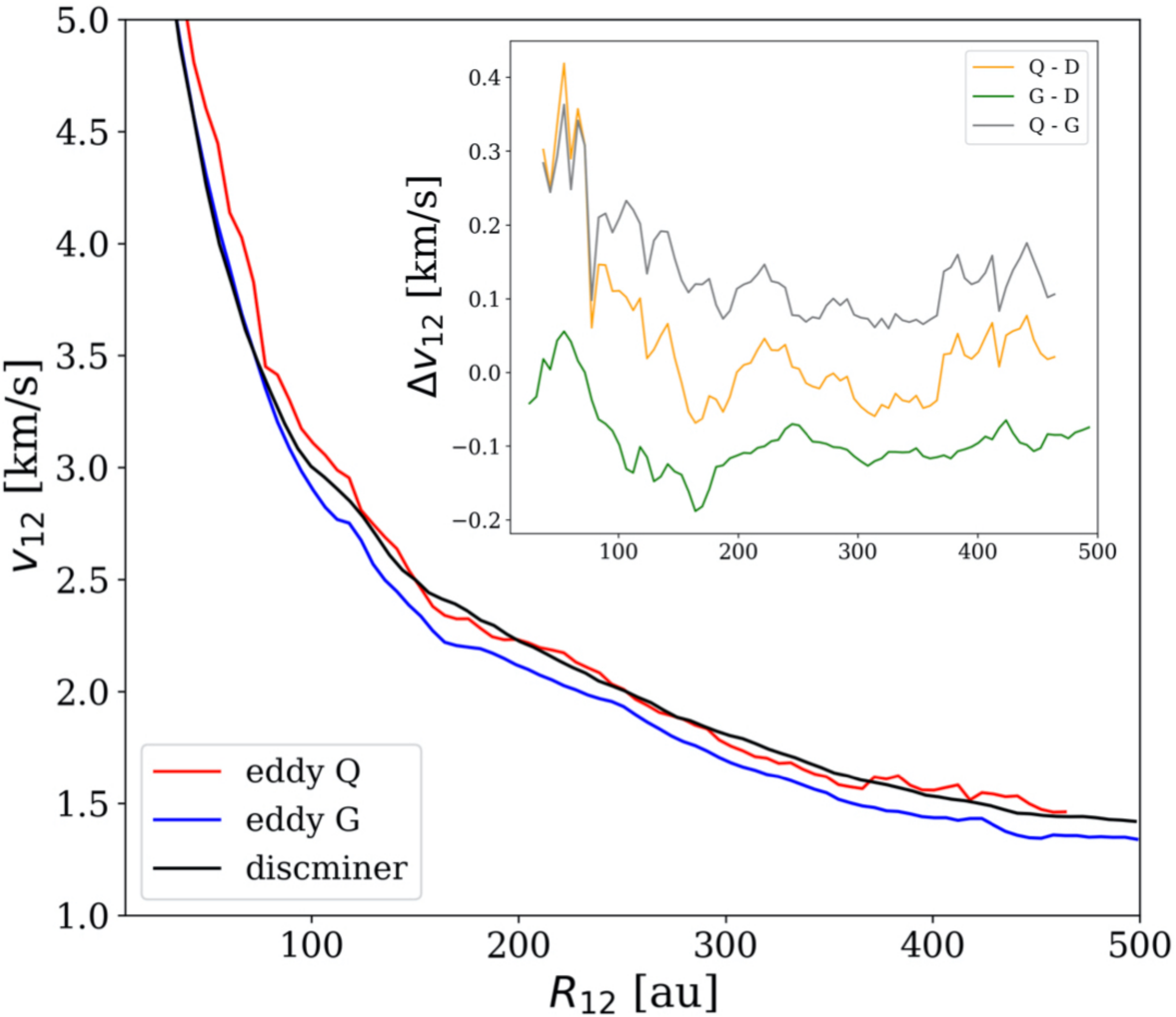}
     \includegraphics[width=\columnwidth]{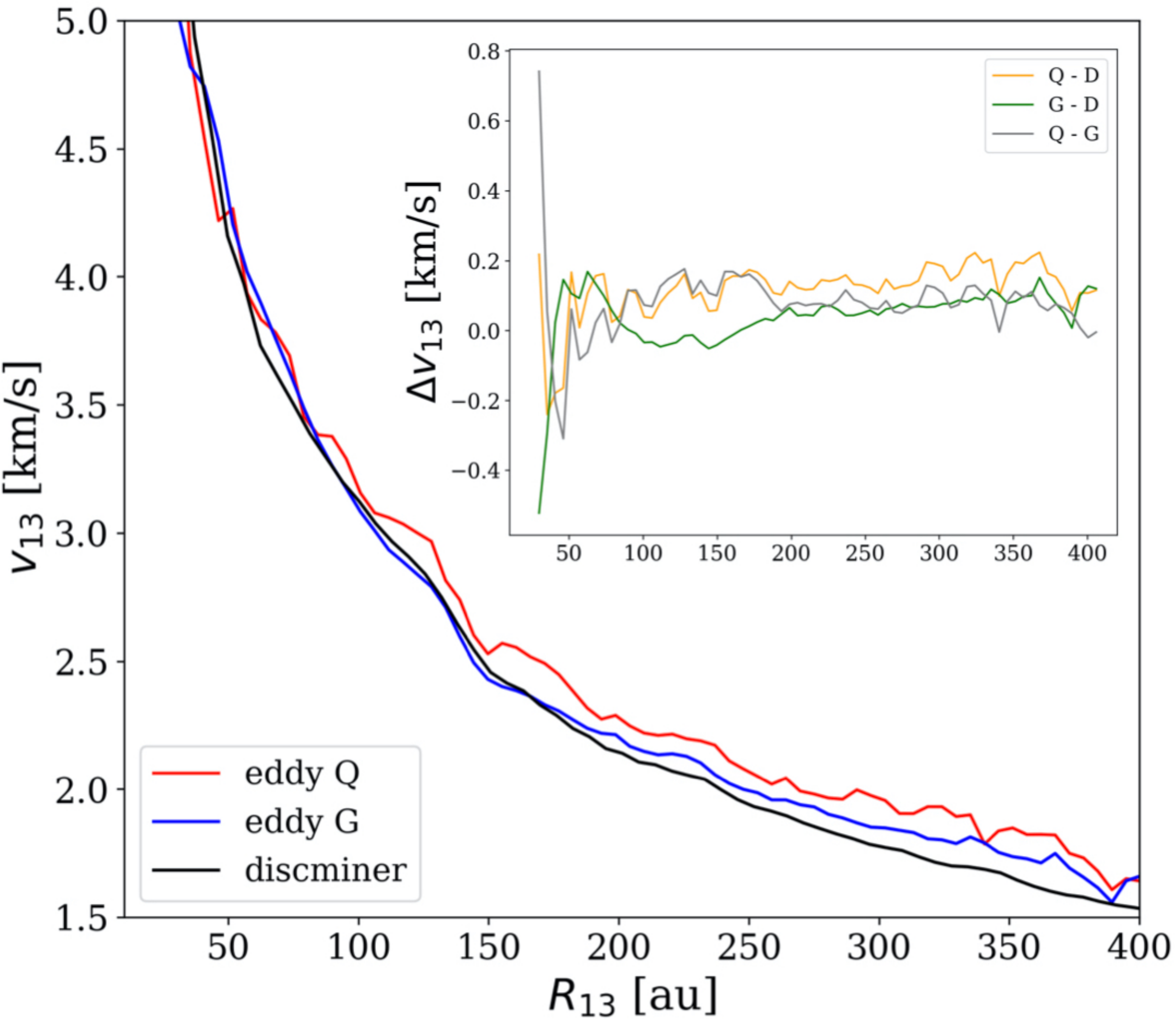} \\
     
     \includegraphics[width=\columnwidth]{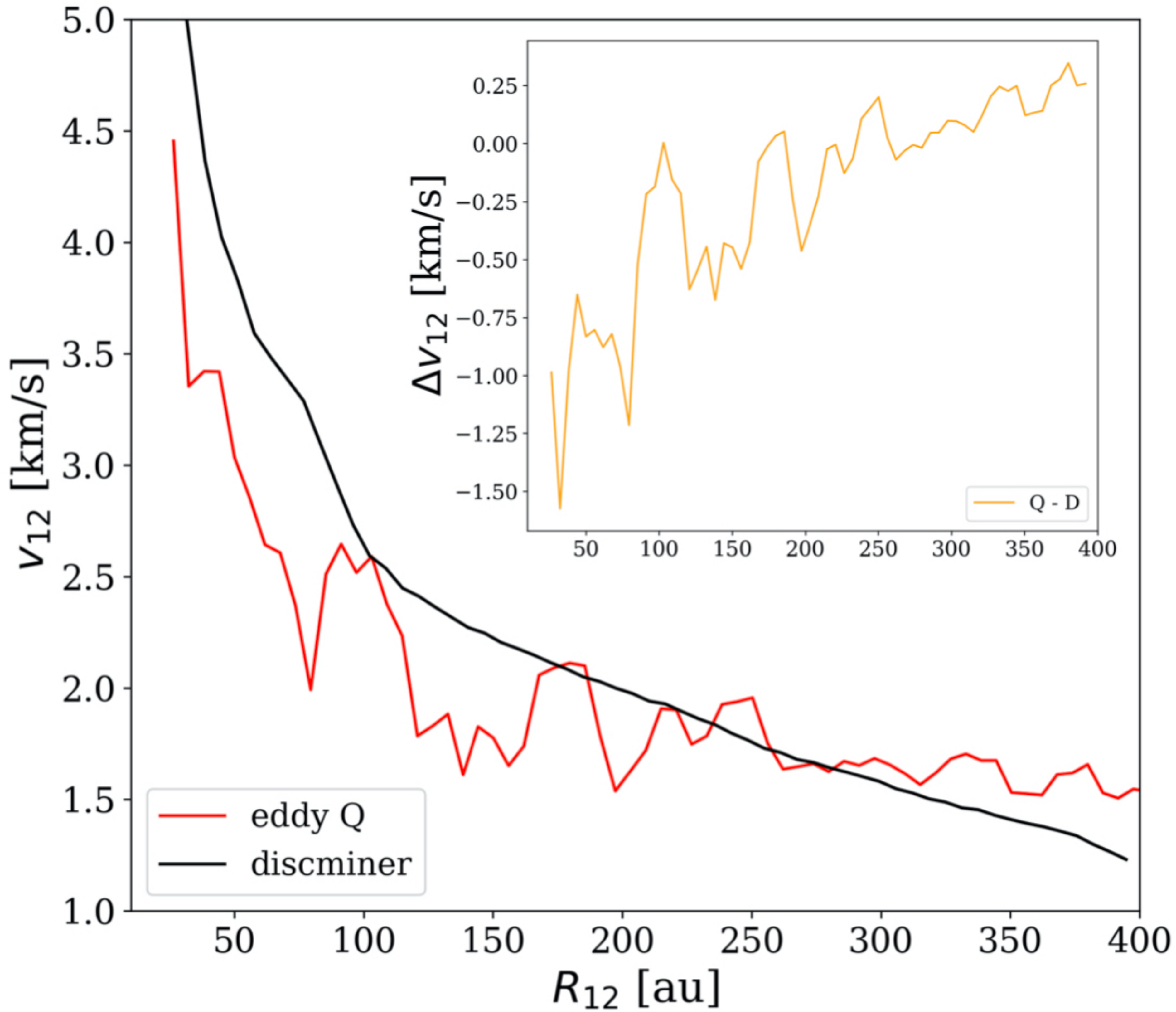}
     \includegraphics[width=\columnwidth]{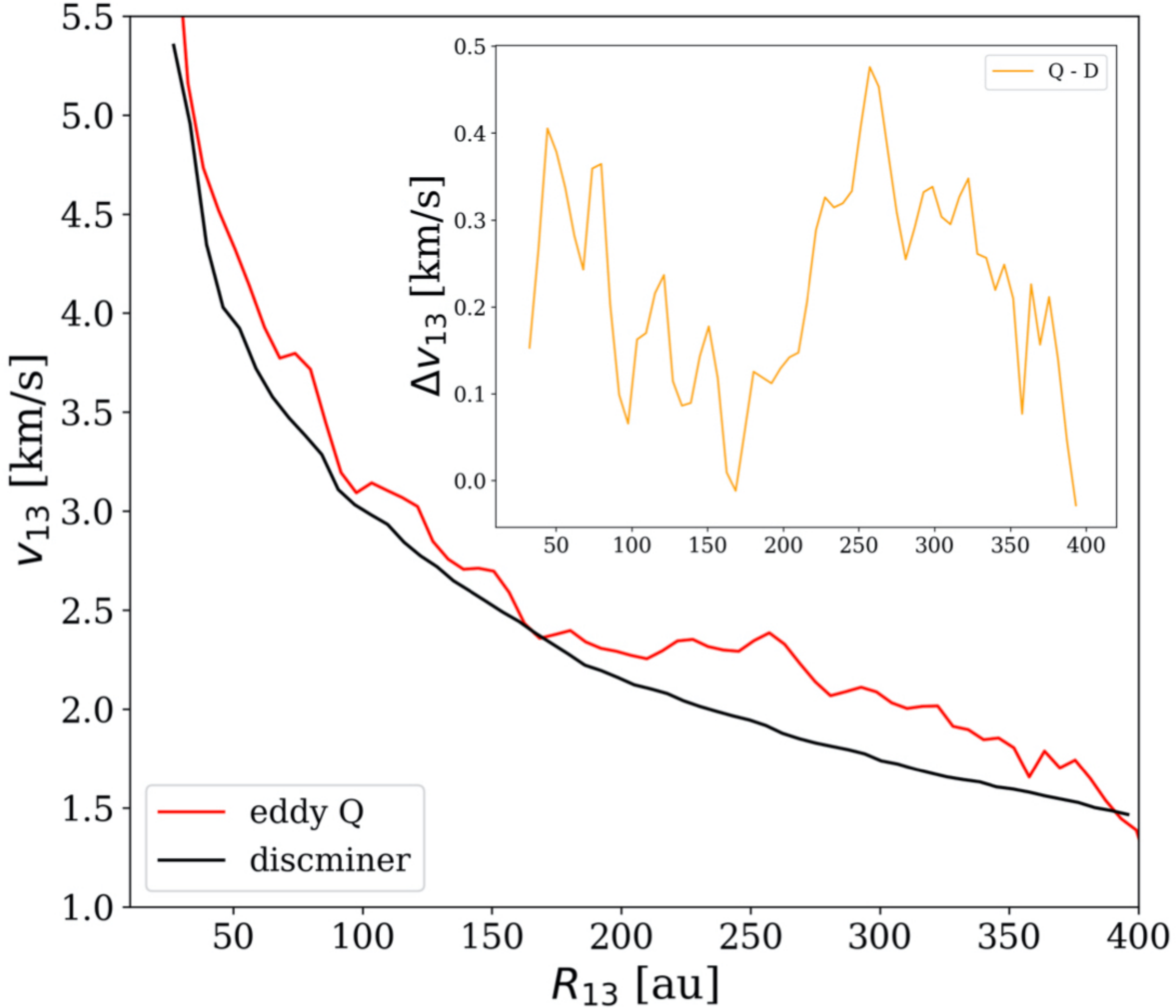} 
    
	\caption{Comparison between the rotation curves extracted with the three different methods: \eddy SHO quadratic (red), \eddy SHO gaussian (blue) and \discminer (black). Upper panels refer to IM Lup (left for the \co line and right for the \col line), lower panels refer to GM Aur (left for the \co line and right for the \col line). {The inset in each panel shows the difference between two curves, where ``Q-D" refers to the difference between \eddy quadratic and \discminer curves, ``G-D" refers to the difference between \eddy gaussian and \discminer curves and ``Q-G" refers to the difference between \eddy quadratic and \eddy gaussian curves}.}

	\label{fig:rotation_curves}
\end{figure*}

\subsubsection{\textsc{EDDY} - Simple Harmonic Oscillator (SHO)}\label{eddy}

To extract the rotation curves of IM Lup and GM Aur using \eddy \citep{teague2019eddy}, we followed the method presented in \citet{teague2018kinematical} and in \citet{teague2018evidence}.

Knowing the height of the emitting layer ($z_i(R)$) for each tracer and the disc geometry (inclination and position angle, see Table \ref{tab:basicparameters}) we can associate the spectrum at any given (projected) location in the disc to its radial distance from the centre of the disc. Considering an axisymmetric disc, we expect the spectra to have the same shape (i.e., peak and width), but to have their line centers shifted by the velocity of the disk, projected along the line of sight ($v_\text{los}$). Under the assumption that the line profiles are the same, we can infer the underlying velocity structure. The easiest approach to take into account the velocity shift is to describe the line centroid as an harmonic oscillator:
\begin{equation}
	v_\text{los}(R,\phi) = v_\phi (R) \sin i \cos \phi + v_\text{sys},
	\label{eq:vlos}
\end{equation}
where $v_\phi$ is the azimuthal velocity (where here we have assumed the motion of the gas as purely azimuthal), $v_\text{sys}$ is the systemic velocity, $i$ is the inclination of the disc, { and $\phi$ is the azimuthal angle in the disk plane. Note that $\phi$ and $R$ are retrieved from any projected location by assuming a thin emitting layer, with height $z(R)$.} There are different methods to fit the line centroids, and in this work we have used the so-called Quadratic method and the Gaussian method, { each of them showing pros and cons. In particular, the Quadratic fit is dependent on the velocity sampling, and it is also sensitive to the channel correlation. As for the Gaussian method, note that the selected velocity range may affect the result in case of skewed profiles}. For a thorough explanation of these methods, see \citet{teague2018robust}. Within this framework, the value of the azimuthal velocity at a fixed radial location is the one that ``aligns'' the spectra (i.e. the location of their peak emission) in the annulus in a velocity-azimuth plot, after shifting the spectrum at each azimuth according to a cosine functional form. More generally, the Simple Harmonic Oscillator (SHO) method implemented in \eddy is able to also obtain the radial velocity by modelling the velocity shift as
\begin{equation}
	v_\text{los}(R,\phi) = v_\phi (R) \sin i  \cos \phi + v_R(R) \sin i \sin\phi + v_\text{sys}.
\end{equation}
To retrieve the azimuthal velocity with the SHO method we need to know the height of the emitting layer of the molecule we are considering, in order to isolate the emission coming from a fixed radial location. { The systemic velocity is fitted as well as the azimuthal and radial ones}.

Yet there is another feature that needs to be accounted for if one is to quantify the rotation velocity of the gas disc correctly. It is known that the impact of the disc lower emission surface on the observed velocities can be critical when it comes to kinematical analyses of high resolution observations of molecular lines in discs (see e.g. \citealt{Izquierdo21,Izquierdo22,Pinte22}). For instance, the lower surface can systematically shift the centroid of the observed intensity profile in an uneven fashion as a function of the disc coordinates, affecting the velocities derived via first moment maps or via parametric fits to the line profile (see e.g. Fig A2 of \citealt{Izquierdo21}). Alternatively, at the cost of velocity accuracy, some methods derive velocities around the peak of the line profile to approximately account for the contribution of the disc upper surface only (see e.g. \citet{teague2018kinematical}). However, when the emission is optically thin, or even marginally optically thick, these methods struggle at distinguishing between the two surfaces as the intensity contrast between both can be very small. When using \eddy, this effect is visible in both sources, and it is particularly relevant for the $^{12}$CO, since its emission layer is higher and the back side of the disc might easily become visible.

We extracted the rotation curves from an inner radius equals to the beam FWHM to an outer radius of $\sim 500$ au for IM Lup in the \co line and $\sim$ 400 au for GM Aur in both lines and IM Lup in the \col line, as the lower surface contribution to the emission becomes dominant in the outer disk.

Some examples of this ``alignment process'' for both IM Lup and GM Aur are shown in Appendix \ref{appendixB}, including both some cases in which the procedure succeeds and some for which it fails.

\subsubsection{DISCMINER}
\label{subsec:discminer}

Similar to the SHO method in \eddy, the extraction of rotation curves with \discminer requires knowledge of the disc orientation and of the height of the molecular emission surface to understand how the disc line-of-sight (l.o.s.) velocities, observed in sky coordinates, translate into the disc reference frame.

The first step of the method consists of obtaining the magnitude of the rotation velocity by deprojecting the l.o.s. velocity retrieved on each pixel by simply inverting Eq. (\ref{eq:vlos}),
\begin{equation}
    v_\phi = \left\langle\frac{(v_{\rm los}(R,\phi)-v_{\rm sys})}{\cos\phi\sin i}\right\rangle_{2\pi}.
    \label{eq:andres}
\end{equation}
This expression assumes that both the radial and vertical components of the velocity field, $v_R$ and $v_z$, are either negligible or localised such that they do not contribute systematically to the observed deviations from Keplerian rotation over large spatial scales. Also \discminer, like \eddy, can obtain the velocity by fitting either a Gaussian or a quadratic form to the line profile.

\textsc{discminer}, however, is able to model and identify the projected location of the disc lower surface to produce azimuthal masks and remove from the analysis those portions of the disc that are more likely affected by the contribution of this surface, which is prominent in the two tracers analysed here for both IM Lup and GM Aur. We find that the deprojected velocities, obtained through Eq. (\ref{eq:andres}), do not suffer from strong azimuthal gradients owing to the contribution of the lower emission surface within an azimuthal section of [-30, 30] deg around the major axes of the discs. It is in this region where azimuthal { unweighted} averages of the deprojected velocities are then computed, as a function of radius, to obtain the rotation curves presented throughout this paper. Further details on this method and quantitative analyses of its accuracy after applying it to synthetic observations of discs with different background velocities will be presented in a separate paper (Izquierdo et al. in prep). { In principle, since in both \textsc{EDDY} and \textsc{DISCMINER} the averages are unweighted, if we restrict \textsc{EDDY} to an azimuthal range $\pm30^\circ$, we should obtain the same curves.}

\subsubsection{Rotation curves: results}

In Figure \ref{fig:rotation_curves} (upper panels) we compare the rotation curves obtained for IM Lup using \eddy (red and grey lines, referring to the different methods to extract the line centroid, see legend), with the one extracted with \discminer (blue line). The differences between the SHO gaussian (red line) and SHO quadratic (grey line) rotation curves are mostly linked to the way each method responds to the contamination of the back side, { and to the velocity sampling and correlation}. Nevertheless, the upper surface emission generally dominates, providing consistent rotation curves. 

For the GM Aur disc (Figure \ref{fig:rotation_curves}, lower panels), the situation is more complicated. The shape of the line centroids in the azimuth-velocity plots is often not well reproduced by a simple cosine function (see also Appendix \ref{appendixB}), resulting in a generally poor fit of the rotation curve. This is due both to contamination of the lower emitting surface and possibly also to a more complex velocity structure in the outer disc (as already discussed in \citet{Huang21}). By using \discminer we are anyway able to provide a good rotation curve, given its ability to model also the lower emitting surface of the disc. 
In general, as evident in Fig. \ref{fig:rotation_curves}, the curves obtained with \eddy are significantly different and much less monotonic that the one obtained with \textsc{discminer}.

In summary, while for the IM Lup disc both \eddy and \discminer provide consistent rotation curves (and the fitted values of disc mass and stellar mass, see Section \ref{sec:results} below, are consistent using both methods), for the more challenging case of GM Aur, the contamination from the back side does not allow to use \eddy and for this source we only use the \discminer rotation curves. The best results are generally achieved by using the Gaussian method for retrieving the velocity, which results in a less bumpy rotation curve. 

\begin{figure*}
\includegraphics[width=0.33\linewidth]{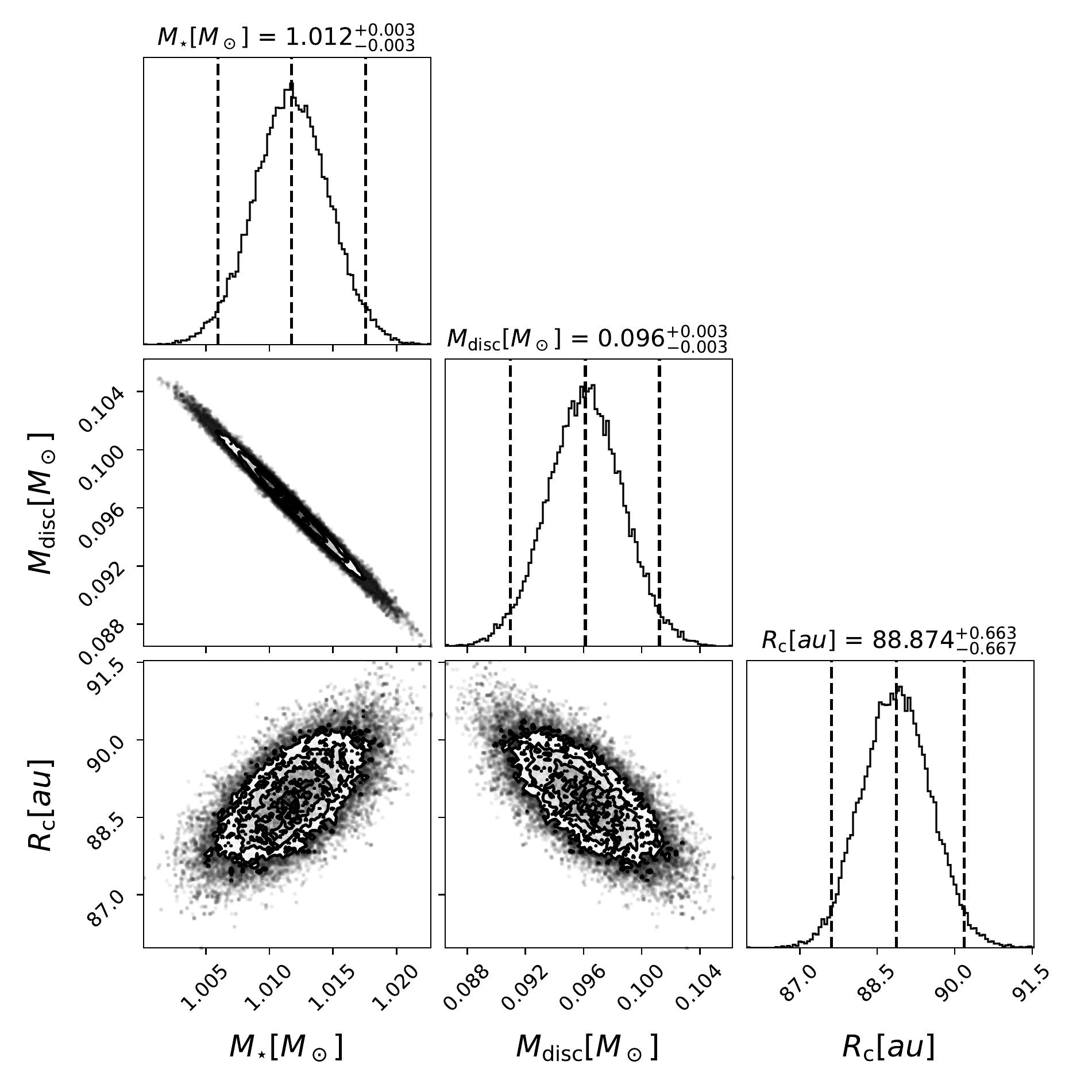}
 \includegraphics[width=0.33\linewidth]{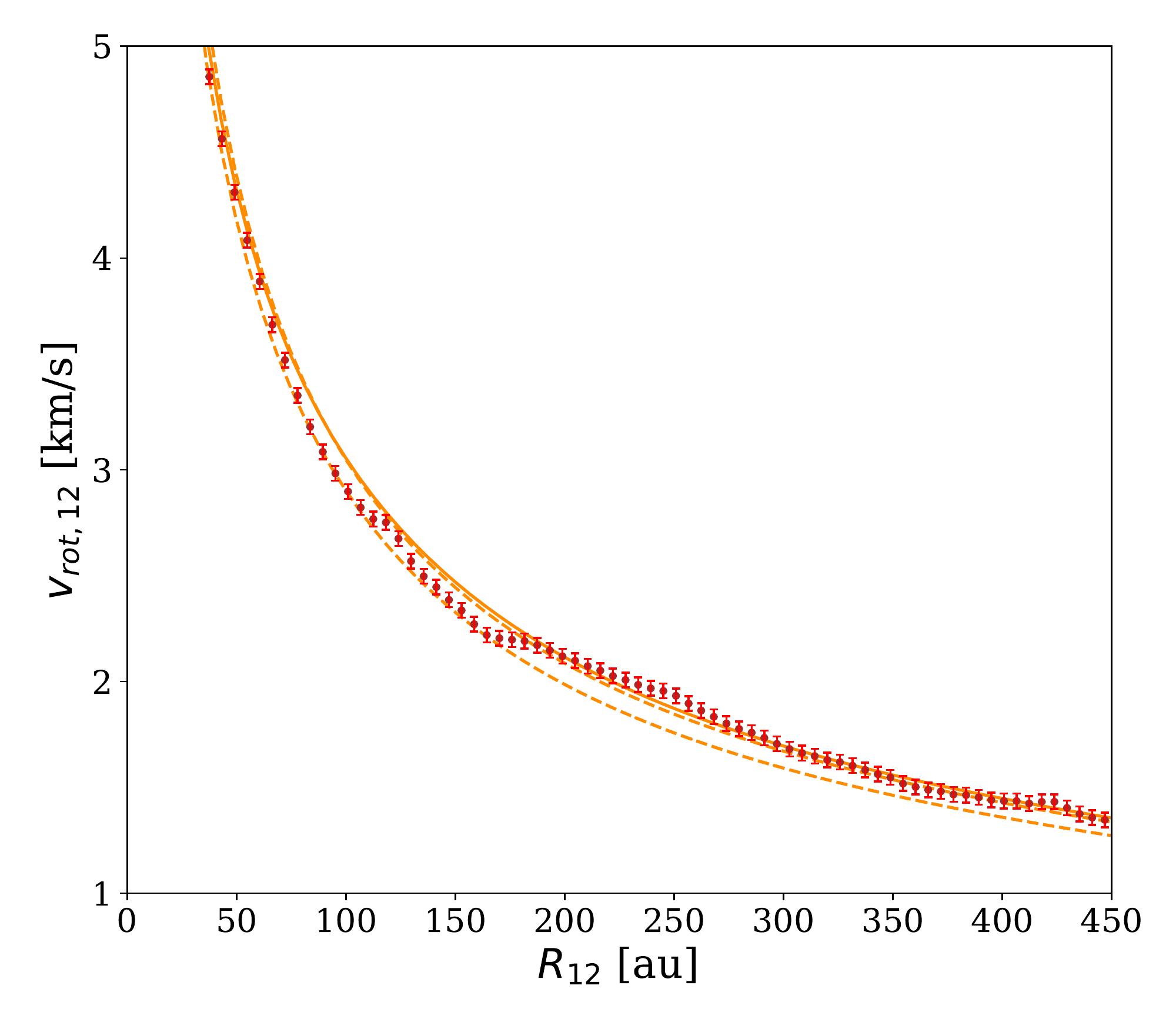}
 \includegraphics[width=0.33\linewidth]{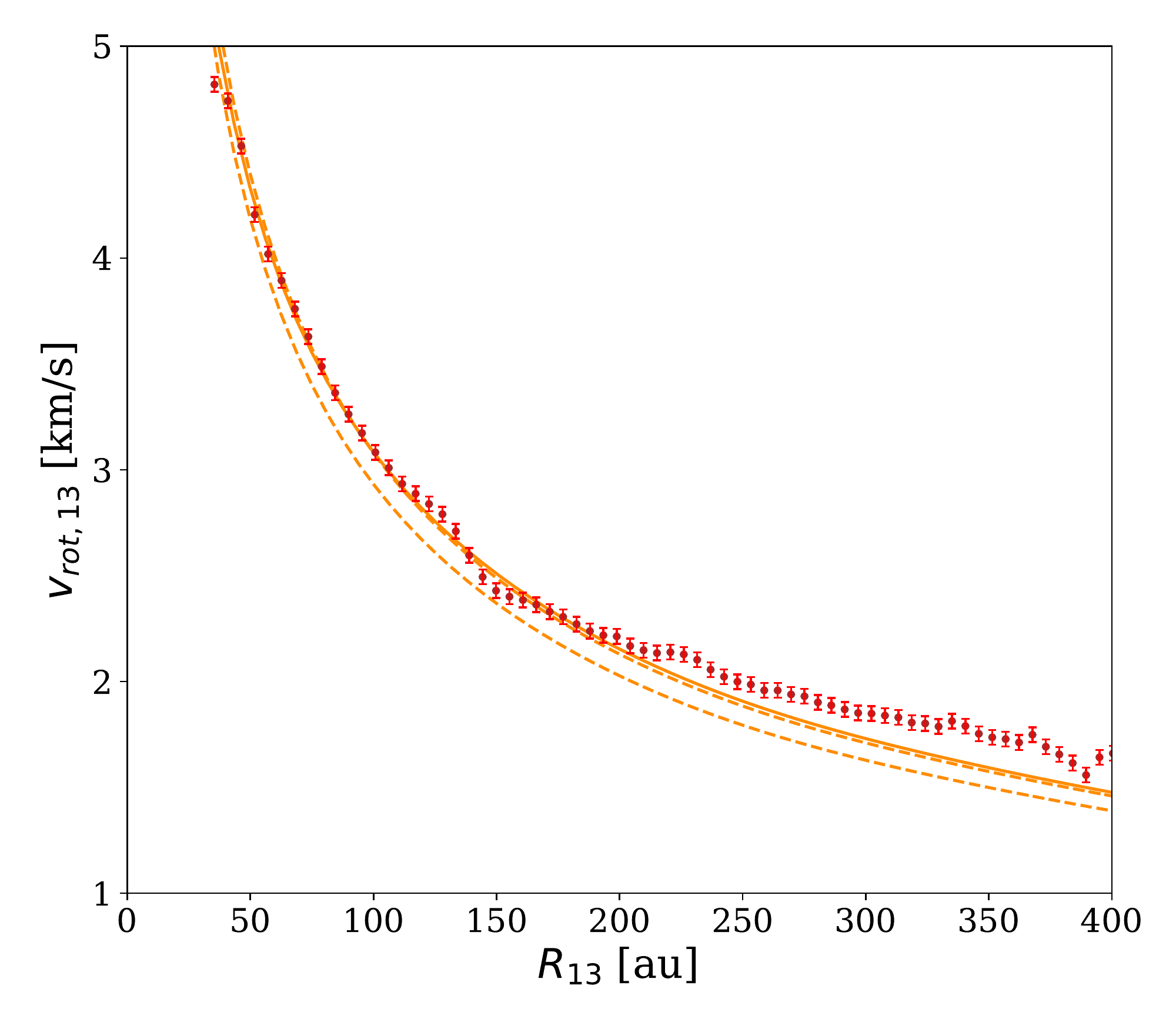}
 \caption{Left: Corner plot of the MCMC fitting procedure showing the distribution of the three relevant fitting parameters: the stellar mass $M_\star$, the disc mass $M_{\rm d}$ and the scale radius of the disc $R_{\rm c}$. Middle: rotation curve of IM Lup obtained from $^{12}$CO data extracted using \eddy, SHO gaussian (red points), along with our best fitting curve including the disc self-gravity (solid line). The lower dashed line indicates the rotation curve obtained by including only the stellar contribution and the pressure gradient, while the upper dashed line is a Keplerian plus pressure curve, where the stellar mass is assumed to be the sum of the disc and star mass from the best fit model. Right: same as the left panel, but for the $^{13}$CO data.}
 \label{fig:rotcurves_im_lup_eddy}
\end{figure*}

\begin{figure*}
\includegraphics[width=0.33\textwidth]{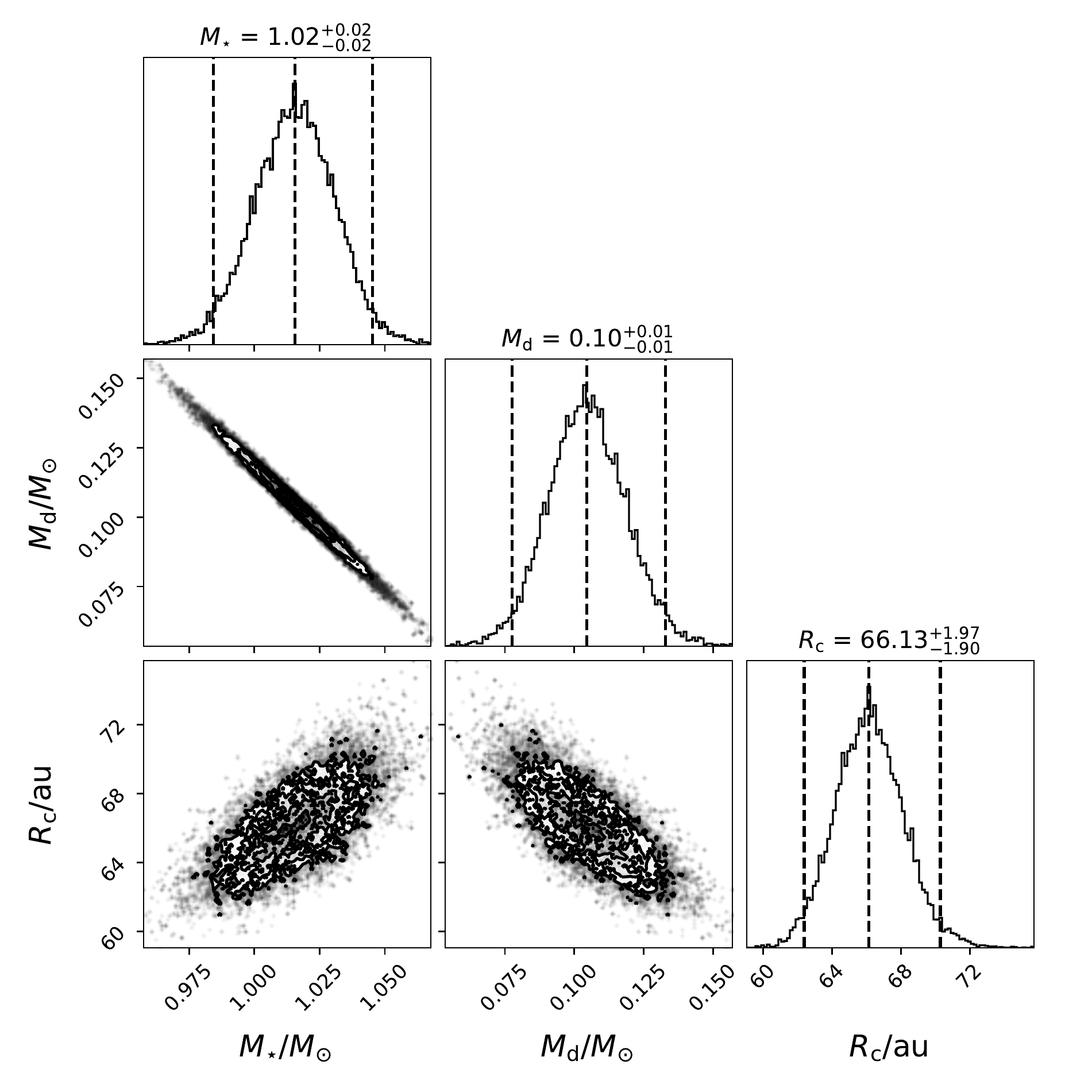}
 \includegraphics[width=0.33\textwidth]{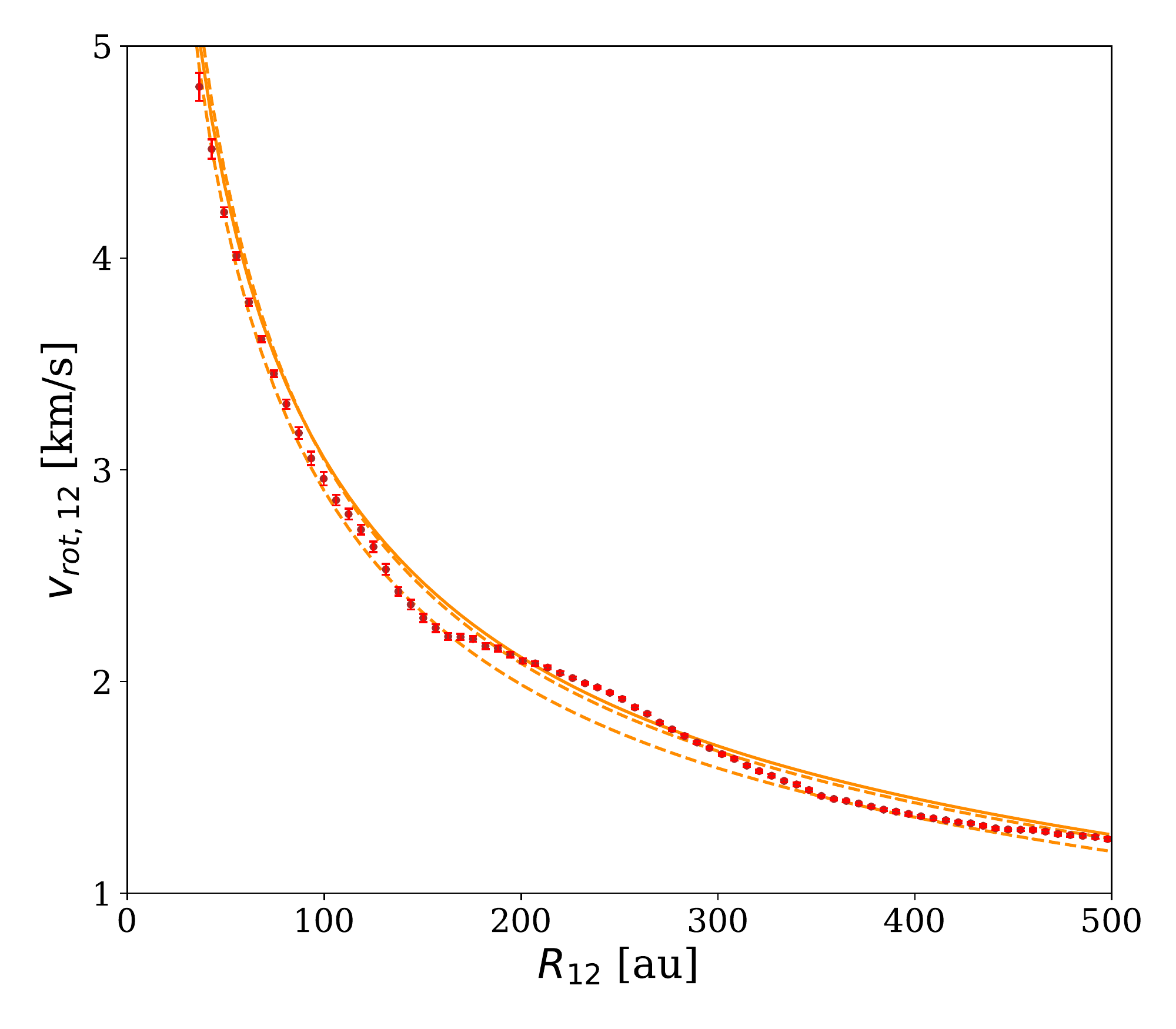}
  \includegraphics[width=0.33\textwidth]{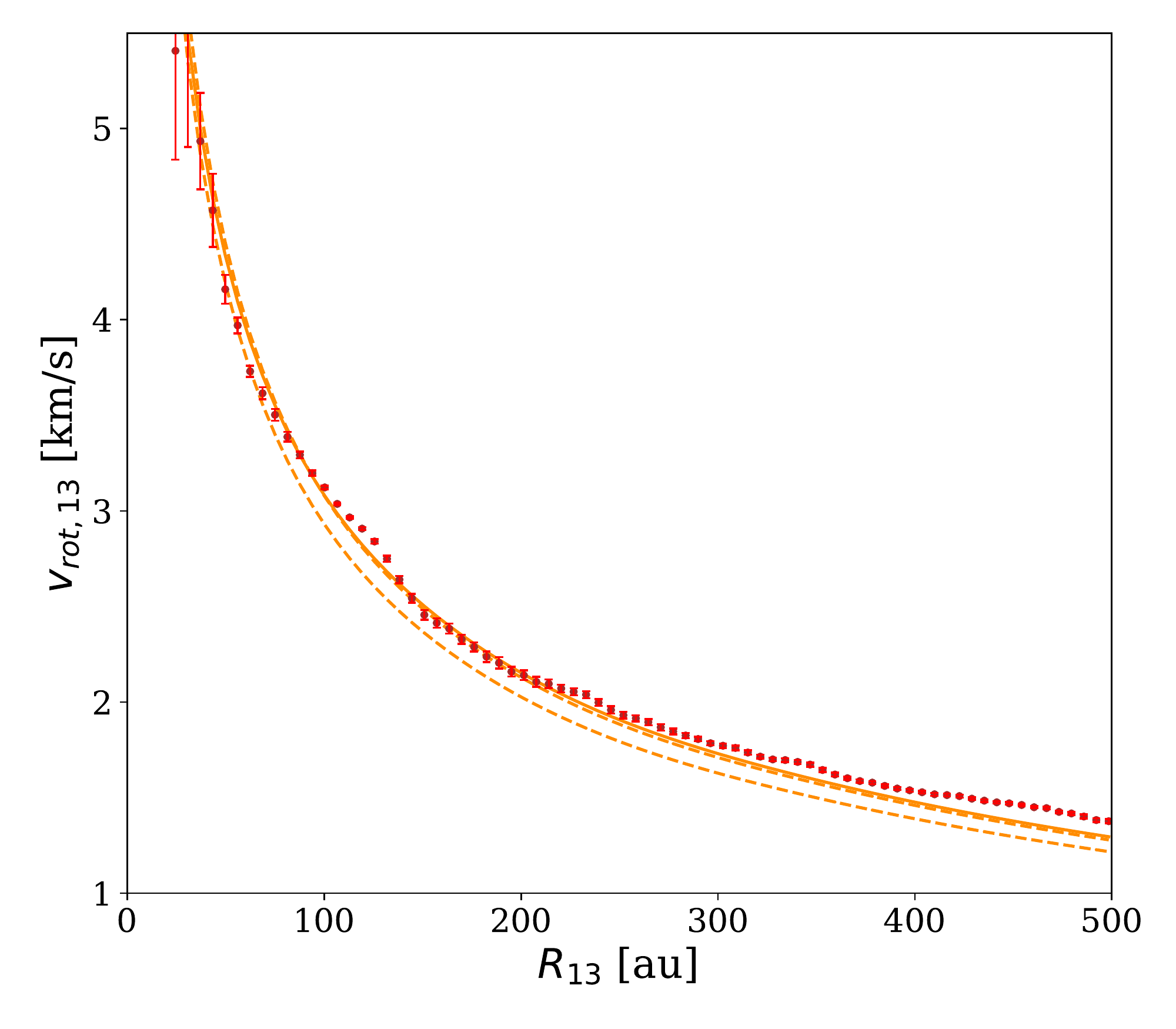}
\caption{Same as Fig. \ref{fig:rotcurves_im_lup_eddy} for IM Lup, but fitting the rotation curves extracted with \discminer (Gaussian).}
 \label{fig:rotcurves_im_lup}
\end{figure*}

\subsection{Modeling the rotation curve}

In centrifugal balance, the rotation of the disc is determined by the following equation:
\begin{equation}\label{rc}
    v_{\rm rot}^2={R\frac{\partial\Phi_{\star}}{\partial R}(R,z)} + R\frac{\partial\Phi_{\rm d}}{\partial R}(R,z) + \frac{R}{\rho}\frac{\partial P}{\partial R}(R,z),
\end{equation}
where $\Phi_{\star}=GM_{\star}/r$ is the stellar potential ($r=\sqrt{R^2+z^2}$ being the spherical radius) and $\Phi_{\rm d}$ is the potential generated by the disc. 

{ Neglecting for the moment the disc contribution to the potential, the rotation curve is given by}
\begin{equation}
    v_{\rm rot}^2  = v_{\rm K}^2 \left\{1-\left[\gamma'+(2-\gamma)\left(\frac{R}{R_{\rm c}}\right)^{2-\gamma}\right]\left(\frac{H}{R}\right)^2 - q\left(1-\frac{1}{\sqrt{1+(z/R)^2}}\right) \right\},
    \label{eq:rotcurve}
\end{equation}
{ and this is valid for a tapered power law surface density, where $v_{\rm K}^2=GM_{\star}/R$ and $\gamma'=\gamma+3/2+q/2$. For a pure power law surface density with exponent $\gamma$, the term $\propto (R/R_c)^{2-\gamma}$ disappears, and we obtain}
\begin{equation}\label{rc_2}
    v_{\rm rot}^2  = v_{\rm K}^2 \left[1-\gamma'\left(\frac{H}{R}\right)^2 - q\left(1-\frac{1}{\sqrt{1+(z/R)^2}}\right) \right],
\end{equation}
{ retrieving the well known result from \citealt{Nelson13} (see also Appendix \ref{appendixA}).}

The disc contribution to the rotation curve is:
\begin{eqnarray}\label{sgterm}
\label{eq:sgpot}
R\frac{\partial \Phi_{\rm d}}{\partial R}(R, z) = G \int_{0}^{\infty} \Bigg[ K(k)-\frac{1}{4}\left(\frac{k^{2}}{1-k^{2}}\right) \times  \\ 
\left(\frac{r}{R}-\frac{R}{r}+\frac{z^{2}}{R r}\right) E(k)\Bigg] \sqrt{\frac{r}{R}} k \Sigma \left(r\right) d r, \nonumber
\end{eqnarray}
{ where $K(k)$ and $E(k)$ are complete elliptic integrals, and $k^2 = 4Rr/[(R + r)^2 + z^2]$ \citep{BL99}. Since $\Sigma$ scales as the disk mass, it is easy to see (cf. Eq. (\ref{eq:sigma})) that the disc contribution to the rotation curve is order $O(M_{\rm d}/M_{\star})$ with respect to the standard Keplerian term. We thus add Eq. (\ref{sgterm}) to the expression of Eq. (\ref{rc_2}) to obtain our model rotation curve of the disc, $v_\text{rot}^2$. We integrate Eq. (14), numerically, as described in Appendix C.}

We see then that there are three corrections to a pure Keplerian profile, $v_{\rm rot}=v_{\rm K}$. The first one is due to the radial pressure gradient, it is sub-Keplerian (i.e. it is a negative contribution to $v_{\rm rot}^2$ and is given by
\begin{equation}
   \frac{\delta v^2_p}{v_{\rm K}^2} = - \left[\gamma'+(2-\gamma)\left(\frac{R}{R_{\rm c}}\right)^{2-\gamma}\right]\left(\frac{H}{R}\right)^2.
\end{equation}
{ This term is generally of the order of $(H/R)^2$, and is important in the outer part of the disc. For typical values of $\gamma^\prime$ and $H/R$, the correction due to the radial pressure gradient becomes important for $R\gtrsim 4R_c$}.

The second one is due to the fact that we evaluate the rotation curve at a finite height $z$ and is due to both the stellar gravitational field and the pressure gradient. It is also sub-Keplerian and is given by 
\begin{equation}
    \frac{\delta v^2_z}{v_{\rm K}^2} = - q\left(1-\frac{1}{\sqrt{1+(z/R)^2}}\right) \approx -\frac{q}{2}\left(\frac{z}{R}\right)^2.
\end{equation}
{Also this term is of order of $(H/R)^2$. Finally, the third correction is due to self-gravity and is provided by Eq. (\ref{eq:sgpot}). The relative importance of these three corrections to Keplerian rotation  depends on the ordering of the two dimensionless parameters $H/R$ and $M_{\rm d}/M_{\star}$. The self-gravity correction is negligible when $M_{\rm d}/M_{\star}\ll (H/R)^2\approx 0.01$. On the other hand, for a marginally gravitationally unstable disc, for which $M_{\rm d}/M_{\star}\approx H/R$, the self-gravity correction dominates over the pressure one\footnote{{ In the marginally unstable case $M_d/M_\star\approx H/R$, and this comes from the fact that the Toomre-Q parameter is equal to 1. Indeed, $Q=H/R / (M_d/M_\star)$.}}. At the same time, there may well be a range of parameters such that $(H/R)^2<M_{\rm d}/M_{\star}<H/R$, in which self-gravity gives a dominant contribution to the rotation curve, while the disc is gravitationally stable \citep{Veronesi21}. { It is worth noting that the pressure-less case, for a power law surface density, had been discussed in the context of AGNs by \cite{hure}}}.

\begin{figure}
    \centering
    \includegraphics[width=\columnwidth]{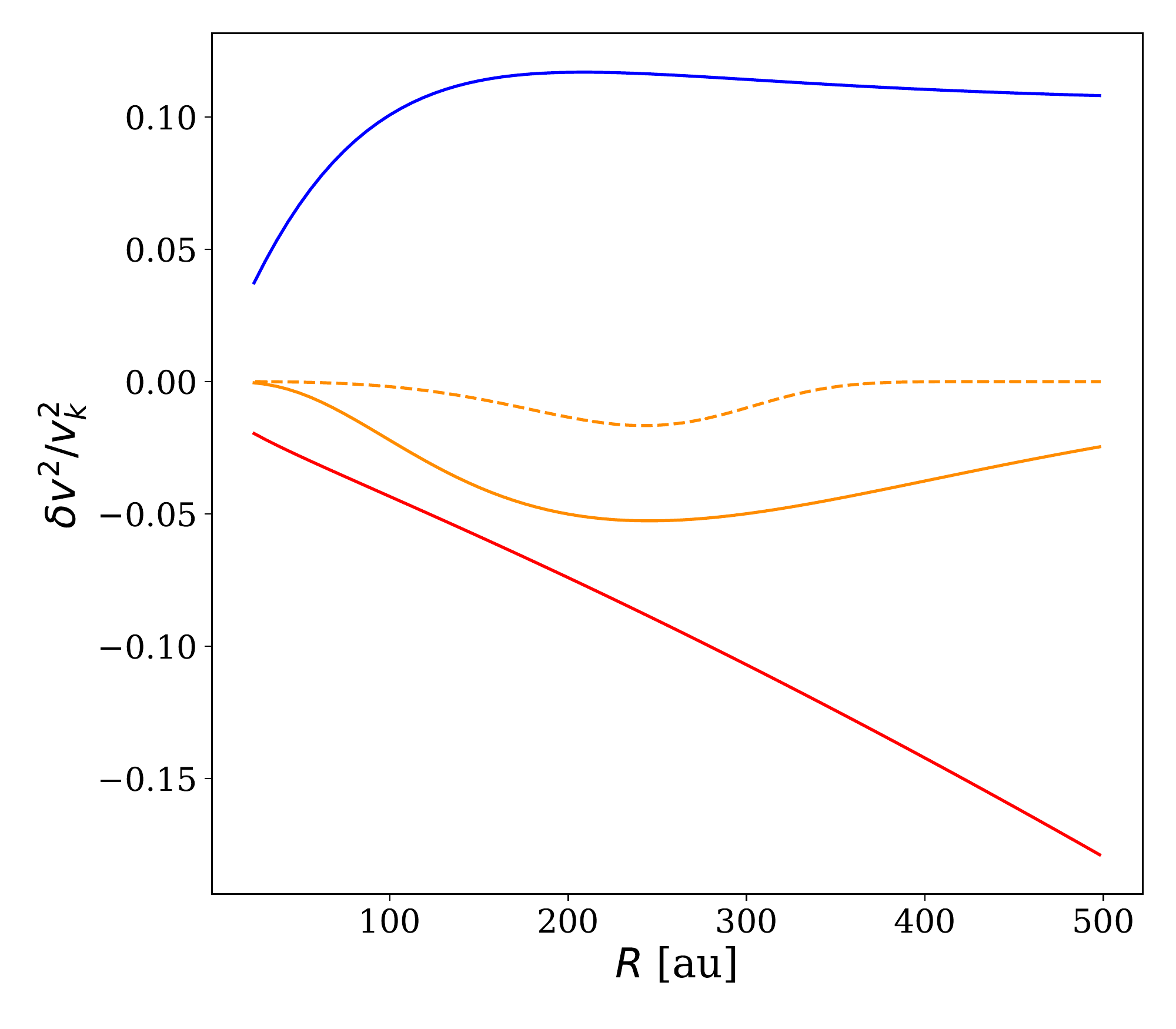}
    \caption{The three sources of non Keplerianity in IM Lup as resulting from our fit. In blue, we show the self-gravitating term, in red the radial pressure gradient $\delta v_p^2$ and in orange the vertical height term $\delta v_z^2$ (solid line for $^{12}$CO and dashed line for $^{13}$CO), emphasising the effect of looking at the curve from a an emission layer at a given height.}
    \label{fig:contributions_im_lup}
\end{figure}
\begin{figure}
    \centering
    \includegraphics[width=\columnwidth]{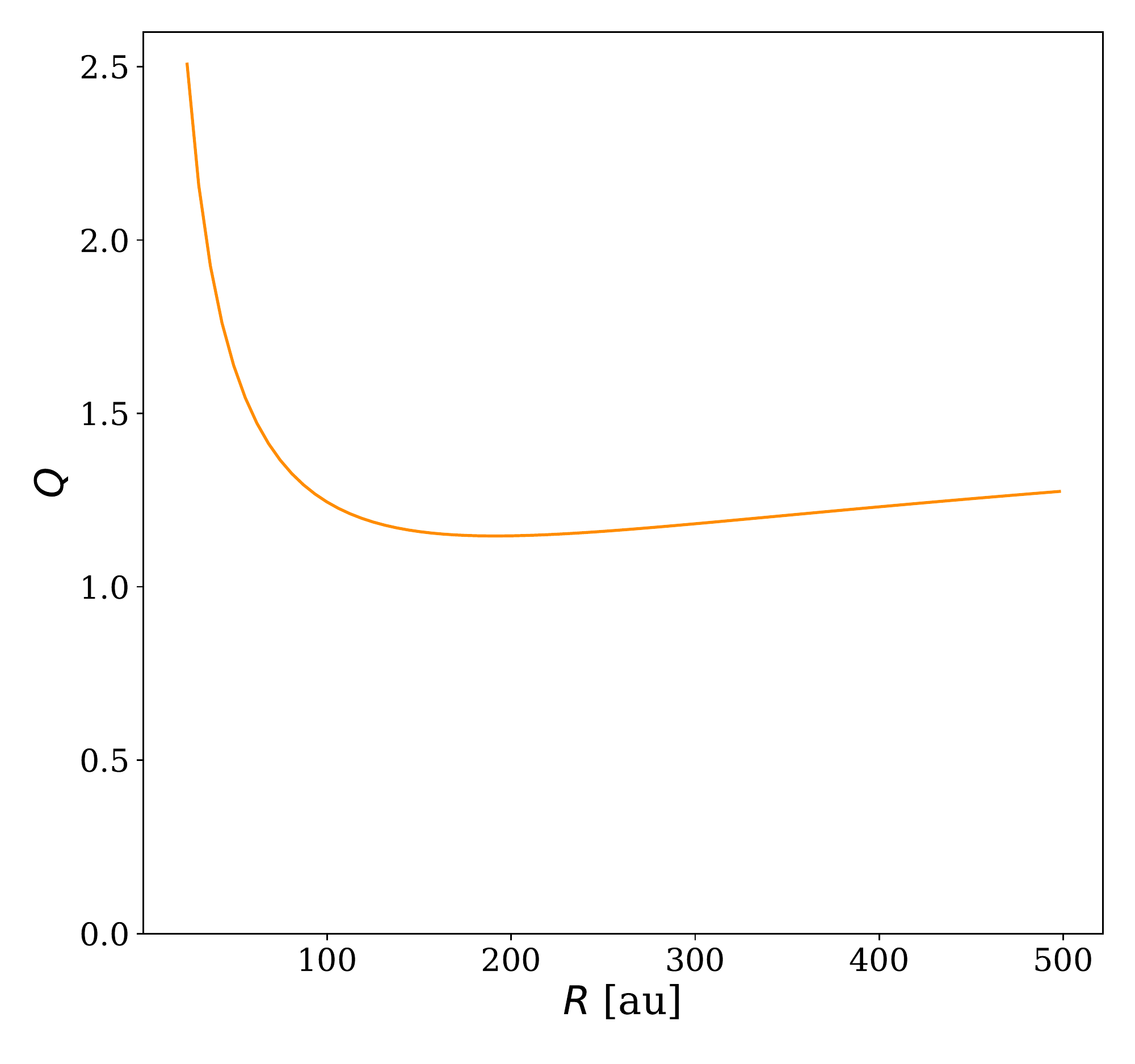}
    \caption{Profile of the stability parameter $Q$ resulting from our best fit model of the rotation curve of IM Lup. At large radii, the { disc approaches $Q\approx 1$}, which might indicate that the disc is marginally unstable.}
    \label{fig:q_im_lup}
\end{figure}
\begin{figure}
    \centering
    \includegraphics[width=\columnwidth]{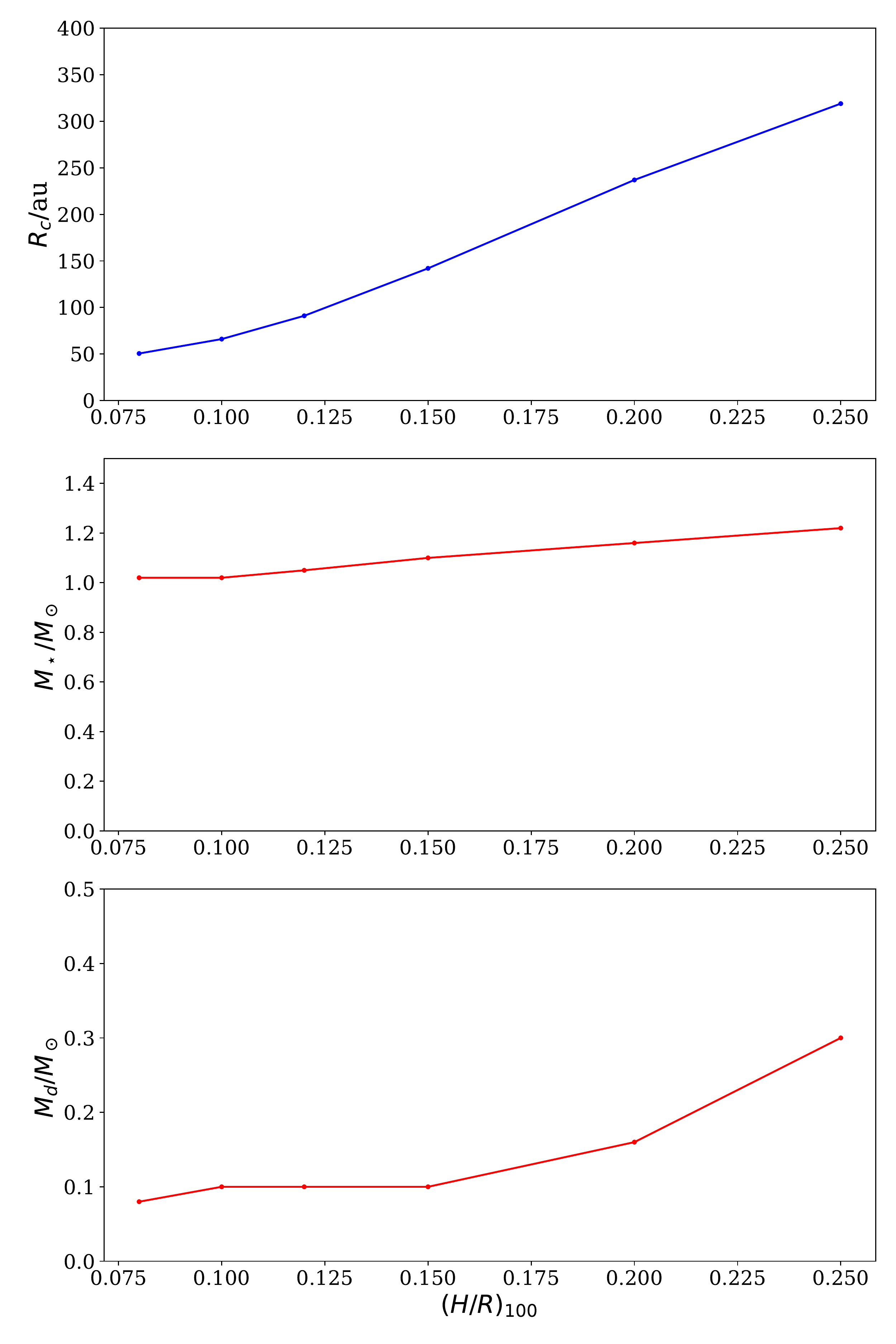}
    \caption{Dependence of the fitted parameters on the assumed value of $H/R$ at $R=100$ au. }
    \label{fig:honr_imlup}
\end{figure}

\section{Results}

\label{sec:results}

Our model rotation curve is given by Eqs. (\ref{eq:rotcurve}) and (\ref{eq:sgpot}) and depends on three free parameters: the stellar mass $M_\star$, the disc mass $M_{\rm d}$ and the scale radius $R_{\rm c}$. In addition, we have to provide the value of $H(R)$ and the height of the emitting layer $z(R)$ for the two isotopologues we consider, as listed in Tab. \ref{tab:basicparameters}. We then fit simultaneously both isotopologues with the same model using standard Monte Carlo Markov Chains techniques, as implemented in emcee \citet{foreman13}. We initialize the MCMC search by assuming a uniform distribution of the three parameters within the following intervals: $M_\star/\text{M}_\odot\in [1,1.5]$, $M_{\rm d}/\text{M}_\odot\in [0.05,1]$, $R_{\rm c}/\mbox{au}\in [100,300]$. We use 100 walkers, with 1500 steps as burn-in and 500 steps to evaluate confidence intervals. The priors are uniform with $M_\star/\text{M}_\odot\in [0,2]$, $M_{\rm d}/\text{M}_\odot>0$, $R_{\rm c}/\mbox{au}\in [50,800]$. { The MCMC fitting procedure code we used is publicly available on GitHub ``DySc - Dynamical Scale"} \footnote{\url{https://github.com/crislong/DySc}.}

\begin{figure*}
\includegraphics[width=0.33\textwidth]{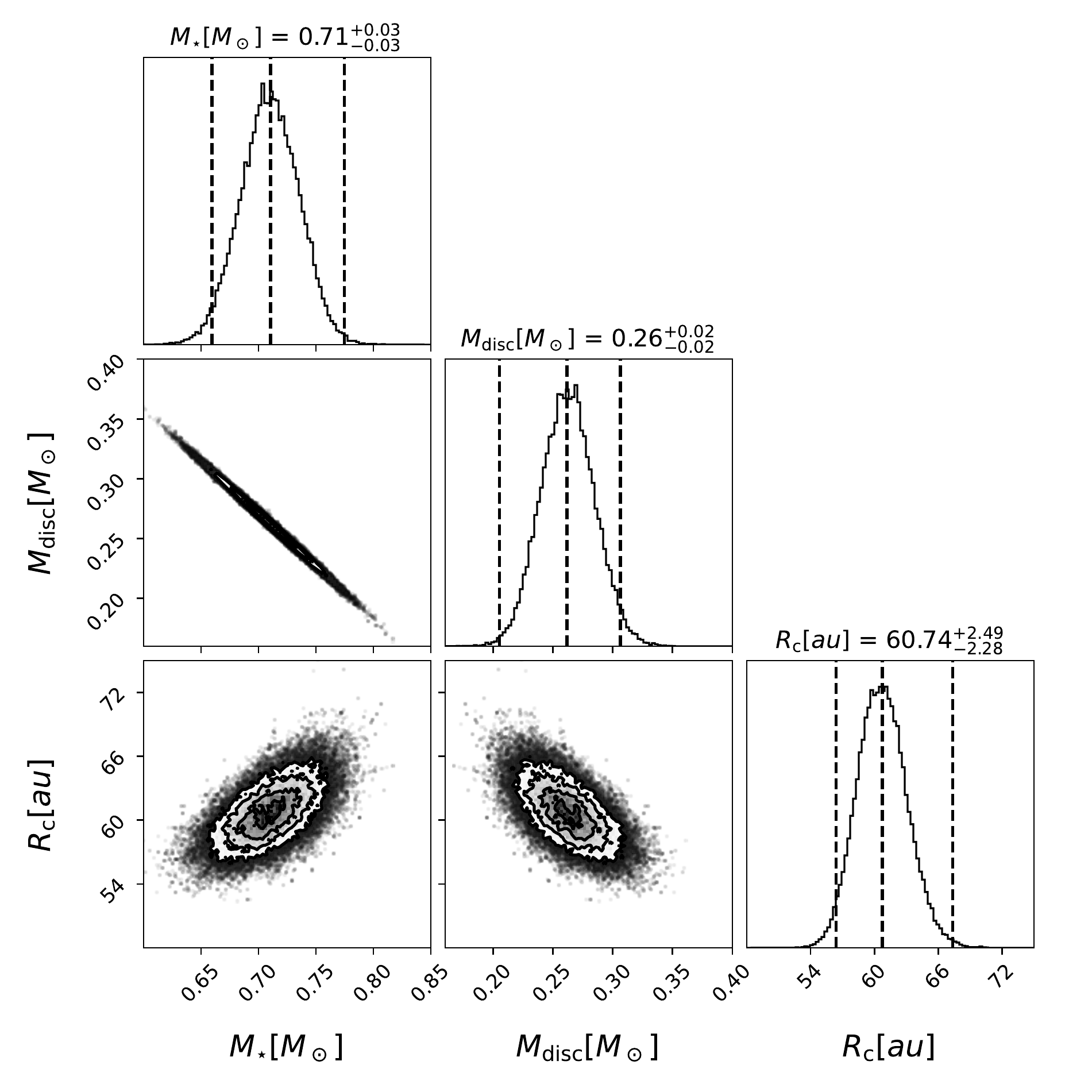}
 \includegraphics[width=0.33\textwidth]{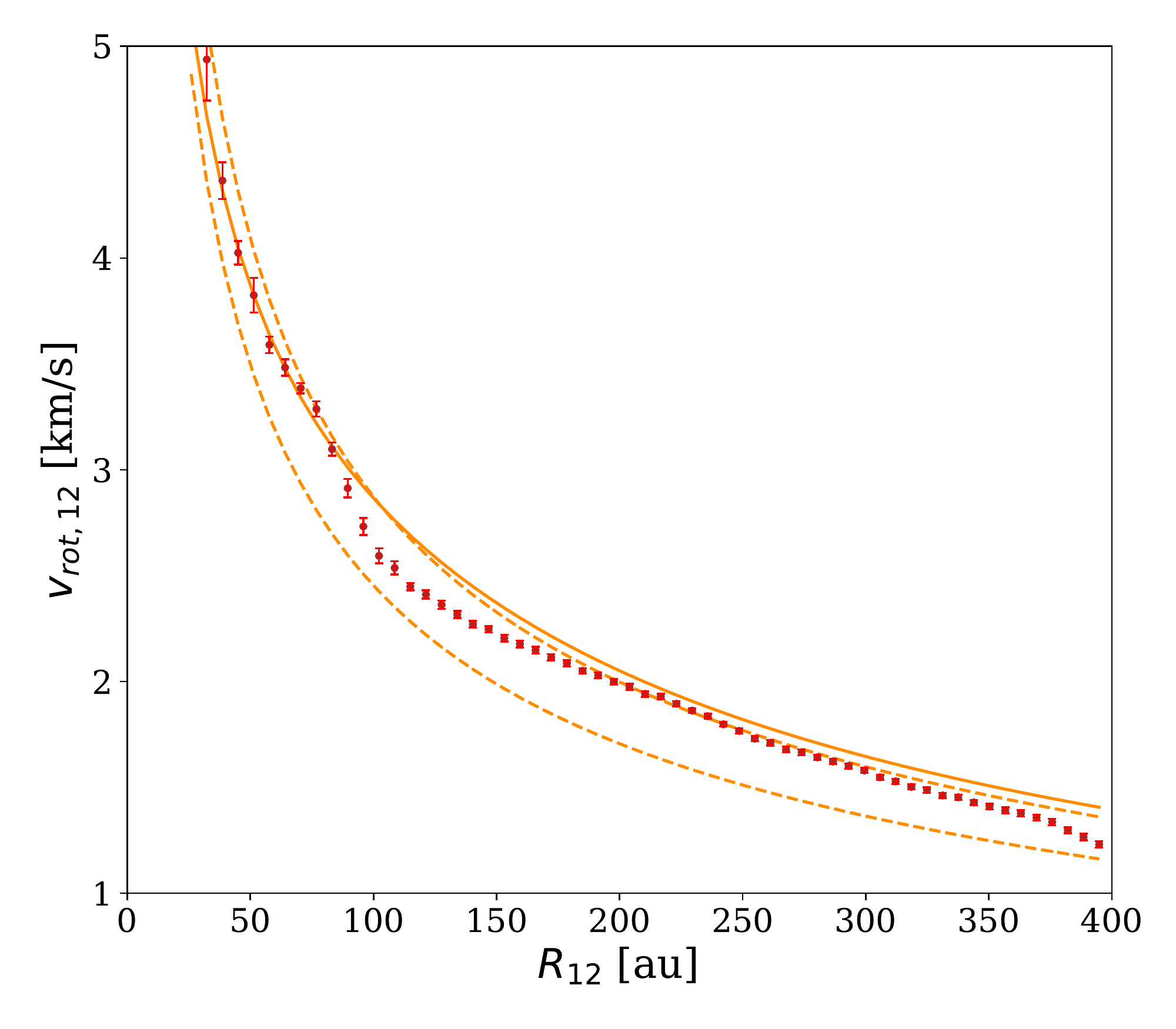}
  \includegraphics[width=0.33\textwidth]{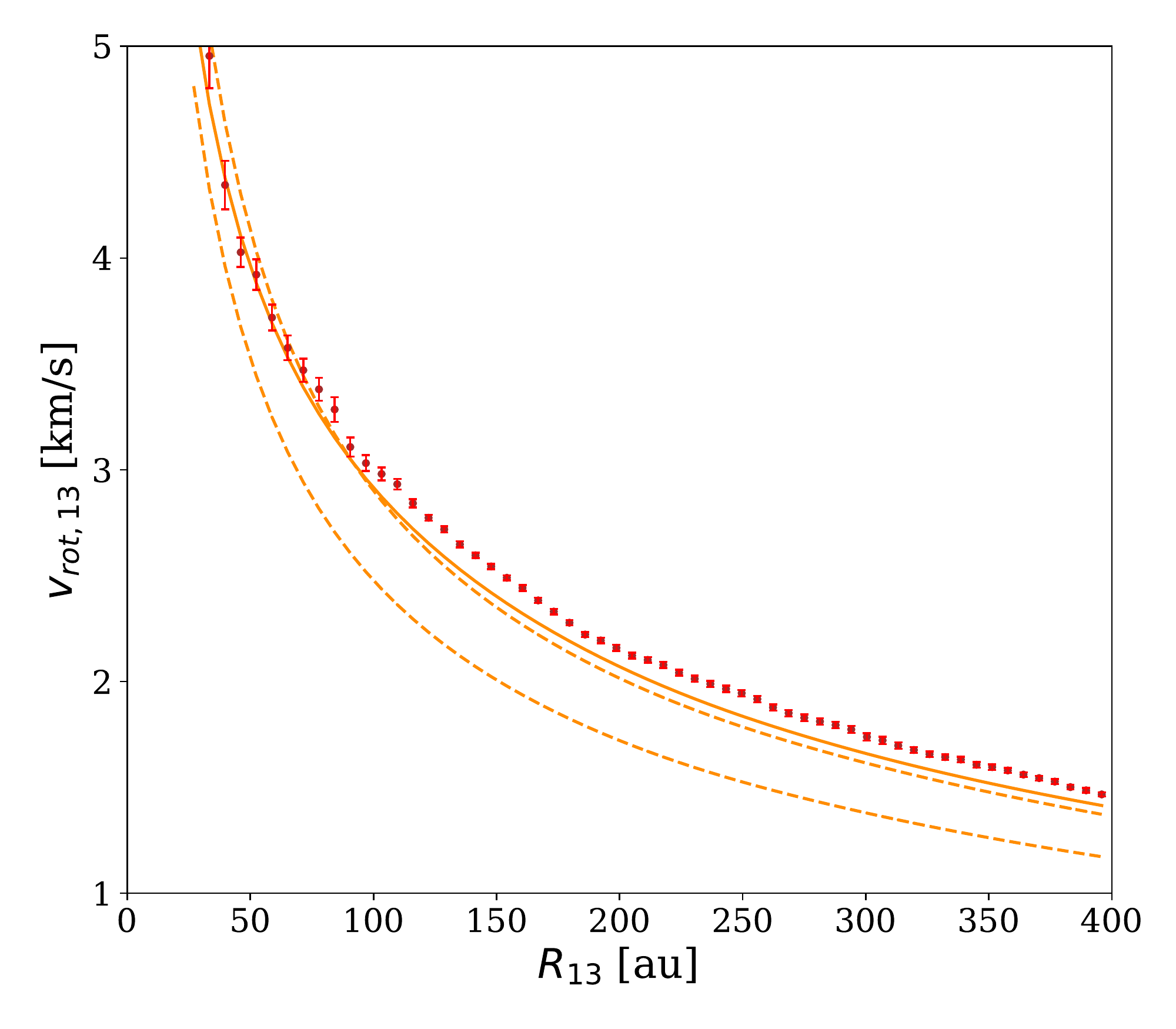}
\caption{Same as Fig. \ref{fig:rotcurves_im_lup}, but for GM Aur.}
 \label{fig:rotcurves_gm_aur}
\end{figure*}

\begin{figure*}
\includegraphics[width=0.35\textwidth]{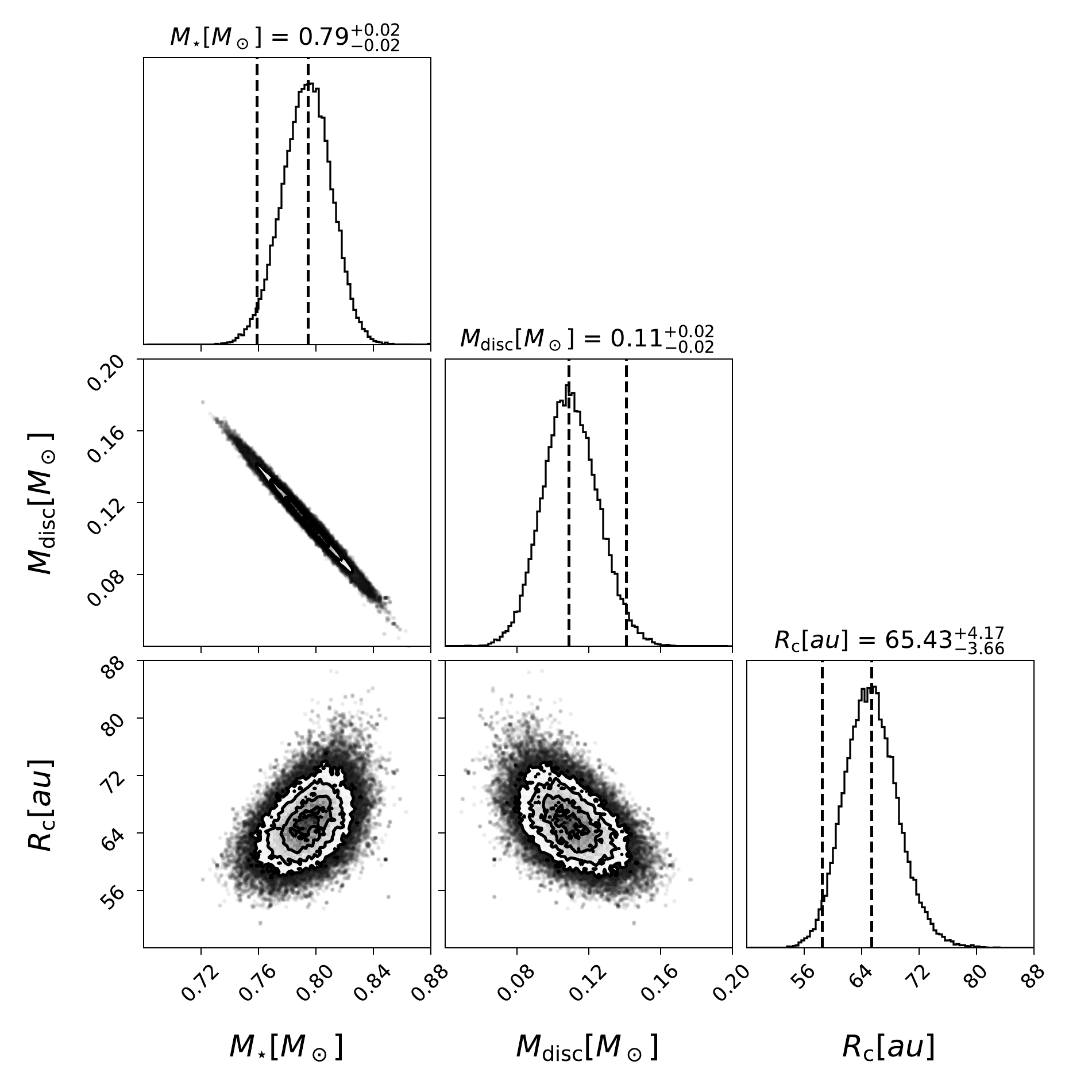}
 \includegraphics[width=0.35\textwidth]{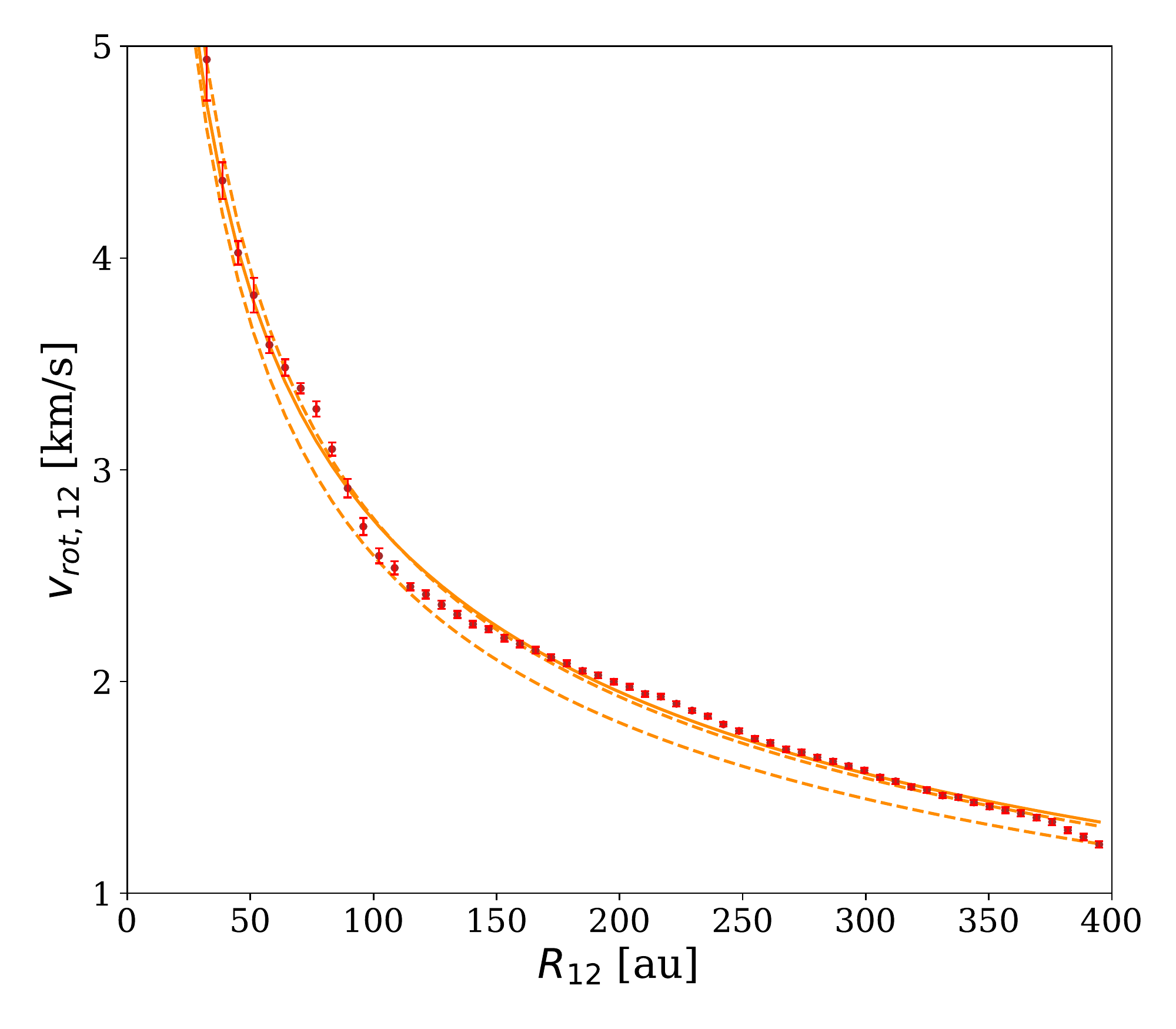}
 
 \includegraphics[width=0.35\textwidth]{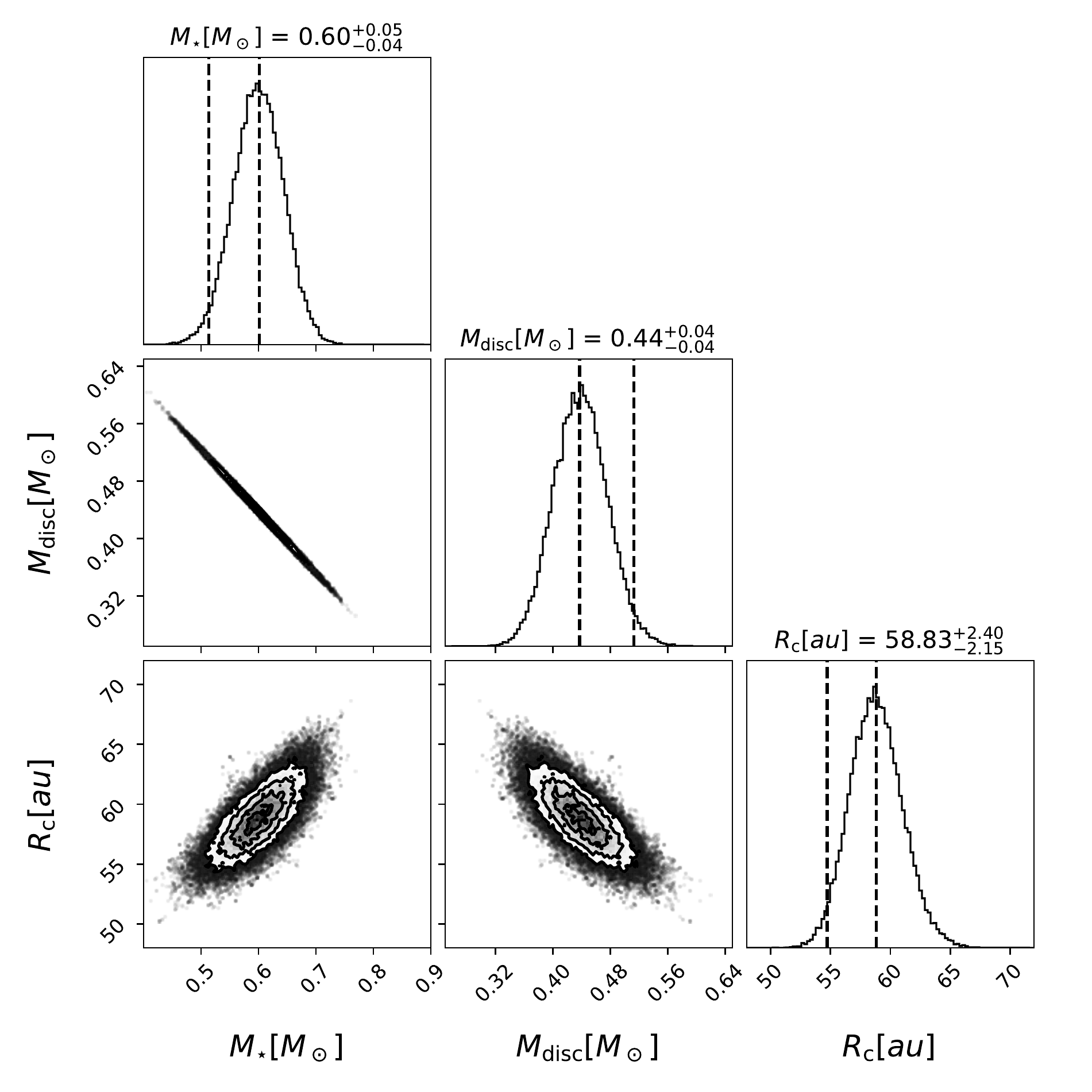}
 \includegraphics[width=0.35\textwidth]{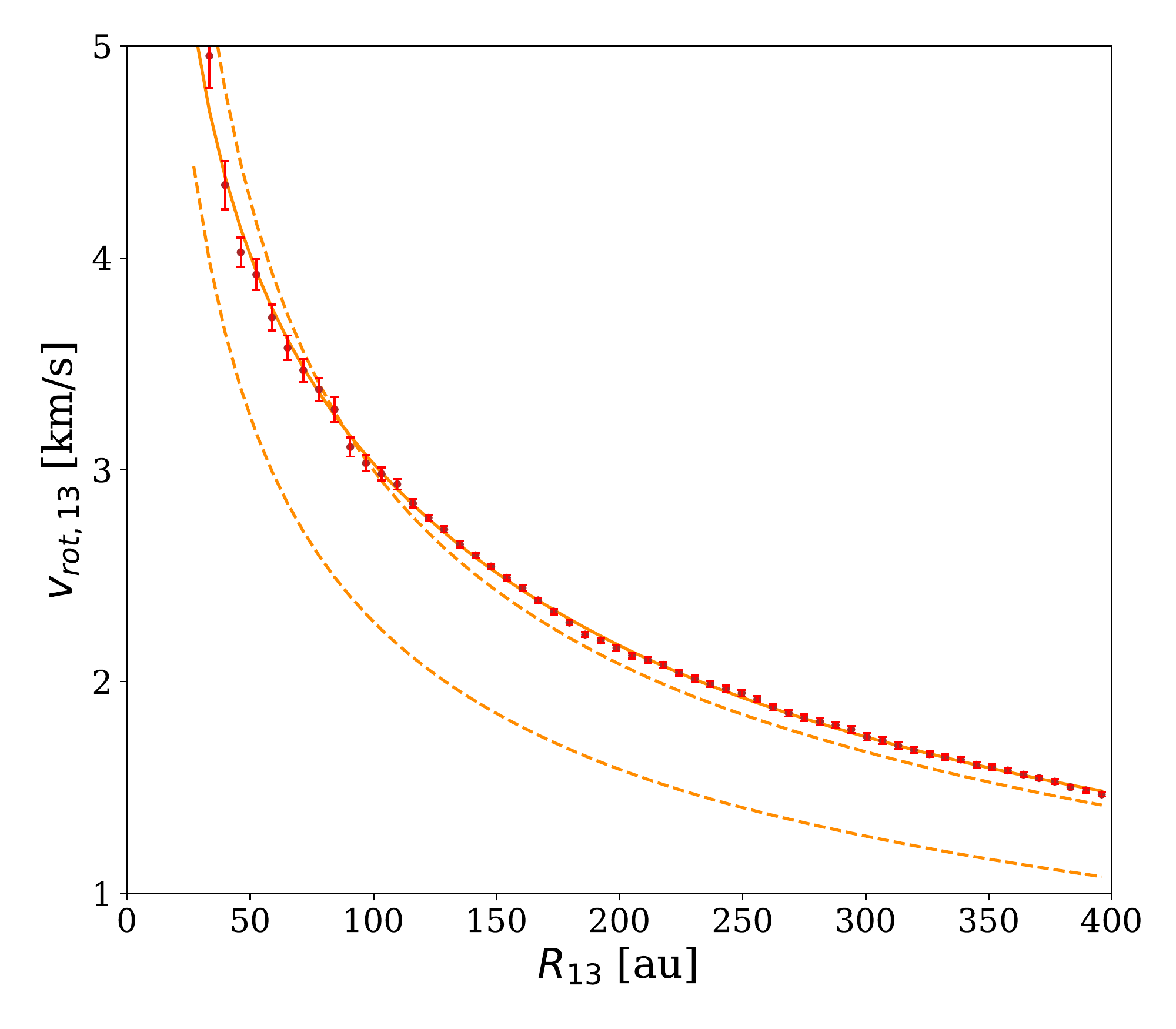}
 
\caption{Upper left: corner plot of our MCMC fitting procedure for the $^{12}$CO rotation curve alone. Upper right: rotation curve of GM Aur obtained from $^{12}$CO data (red points), along with our best fitting curve including the disc self-gravity (solid line). The lower and upper dashed line indicate the rotation curves obtained as described in Fig. \ref{fig:rotcurves_im_lup}. Lower left: corresponding corner plot for the $^{13}$CO line. Lower right: corresponding data and best fit rotation curves for the $^{13}$CO line.}
 \label{fig:gmaur_12}
\end{figure*}

\begin{figure}
    \centering
    \includegraphics[width=\columnwidth]{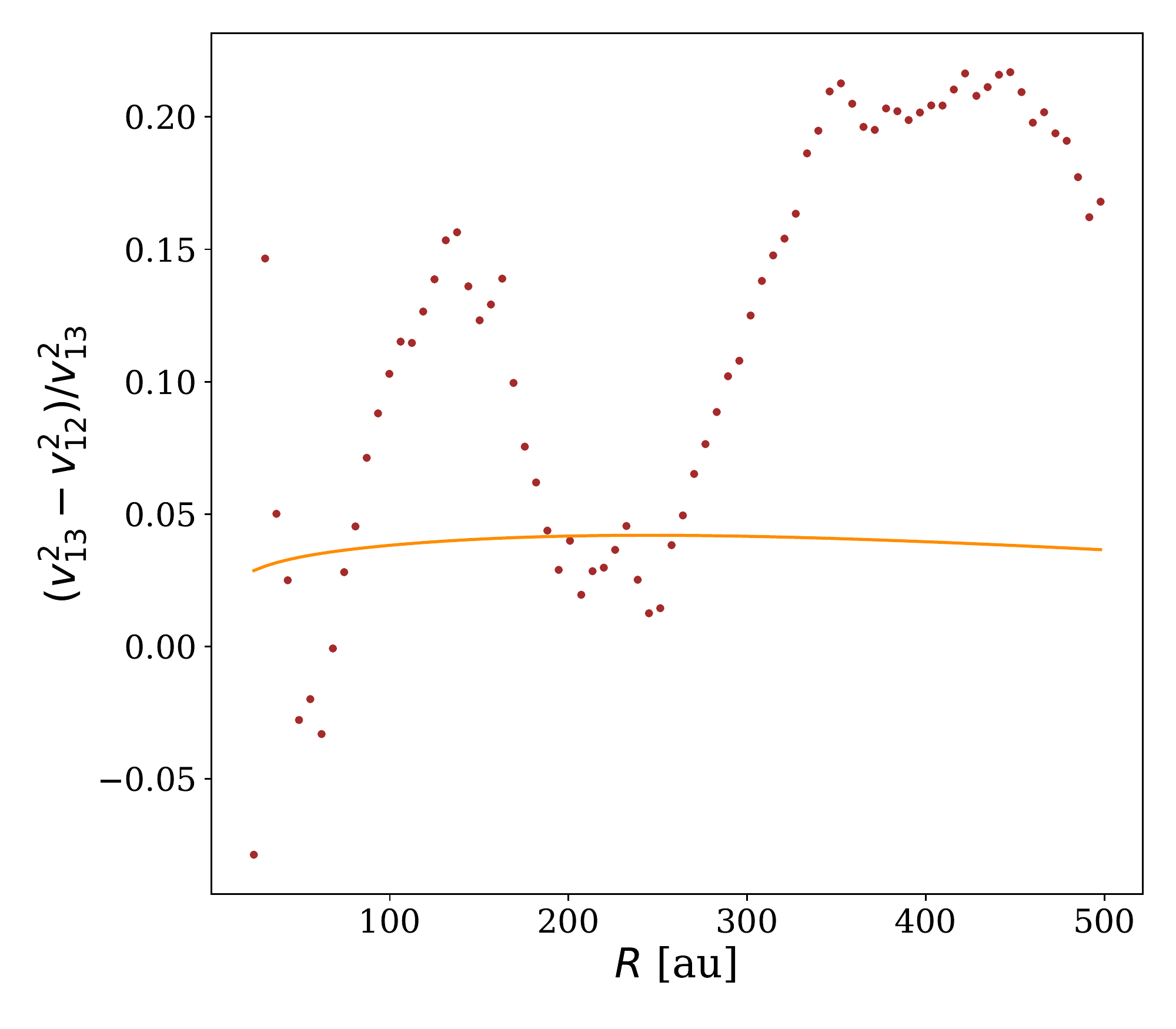}
    \caption{Red dots indicate the observed difference between the rotational velocity of $^{13}$CO and $^{12}$CO as obtained from the \discminer, while the orange line shows the expected difference due to vertical stratification, assuming the temperature structure of the two molecules as described in \citet{Law21}.}
    \label{fig:dv}
\end{figure}

\subsection{IM Lup}

Figures \ref{fig:rotcurves_im_lup_eddy} and \ref{fig:rotcurves_im_lup} show the results of our fitting procedure to the rotation curves of IM Lup, obtained with \eddy and \textsc{discminer}, respectively, using the Gaussian method. The two methods appear to be in very good agreement between themselves.
For \eddy, the best fit stellar mass is $M_\star\approx 1.02 \, \mathrm{M}_{\sun}$, the best fit disc mass is $M_{\rm d}\approx 0.096 \, \mathrm{M}_{\odot}$ and the scale radius is $R_{\rm c}\approx 66$ au, as shown in the corner plot in Fig. \ref{fig:rotcurves_im_lup_eddy} (left panel). For \discminer, the best fit stellar mass is $M_\star\approx 1.01 \, \mathrm{M}_{\sun}$, the best fit disc mass is $M_{\rm d}\approx 0.11 \, \mathrm{M}_{\odot}$ and the scale radius is $R_{\rm c}\approx 67$ au, as shown in the corner plot in Fig. \ref{fig:rotcurves_im_lup} (left panel).
The corresponding rotation curves are shown in Figs. \ref{fig:rotcurves_im_lup_eddy} and \ref{fig:rotcurves_im_lup} (middle panel: $^{12}$CO, right panel: $^{13}$CO), where the red dots indicate the data points, while the three curves show our best fitting model including the disc self-gravity (solid line), a model rotation curve including only the stellar contribution to the potential, and a model curve including only the stellar contribution, but with a mass equal to the sum of the disc and stellar mass in our best fitting model (dashed lines). As one can see, the data points transition from an inner curve which is well represented by the stellar potential to an outer region where the curve is approximated by a Keplerian curve, but with a higher mass including both contributions to gravity. 

{ Note that in the fitting procedure we have neglected the fact that the measurement points of the rotation curve are not fully independent. This has an impact on the estimated parameter uncertainty (here and for all the other fits presented below), which is in fact an underestimate. To prevent this issue, it would be necessary to model the covariance matrix of the system, which is beyond the scope of the present work. Moreover, note that these uncertainties only takes into account the statistical errors and not the systematics associated with the derivation of the rotation curve or with the other assumptions on the geometrical parameters of the disc, that are discussed below.}

In Figure \ref{fig:contributions_im_lup} we show the three individual corrections to Keplerianity as resulting from our fit. The term arising from the height of the emission layer is generally sub-dominant, contributing at most to $\sim 5$\% of the Keplerian term. On the other hand, the pressure term and the self-gravity term are comparable but opposite in sign, reaching a fraction of $\sim 10-15$\% of the Keplerian velocity in the outer disc.

Figure \ref{fig:q_im_lup} shows the profile of the $Q$ parameter,
\begin{equation}
    Q=\frac{c_{\rm s}\kappa}{\pi G \Sigma}\approx \frac{H/R}{M_{\rm d}(R)/M_\star},
\end{equation}
defining the stability of the disc against gravitational instabilities resulting from our best fit model, where $\kappa\approx \Omega$ is the epicyclic frequency. In the outer disc, our model predicts $Q\sim 1$ indicating that the disc might be marginally gravitationally unstable, consistent with the observed large scale spiral structure in the dust morphology \citep{Huang18}. 

Note that, given the known value of the dust mass of $M_{\rm dust}=1.7~10^{-3}\, \mathrm{M}_{\sun}$ \citep{Cleeves16}, our measurement implies a gas/dust ratio of $\sim 60$, close to the standard assumed value of 100. 

\subsubsection{Dependence on $H/R$}

In our reference model, described in the section above, we fix the pressure scale height to the literature value $H_0=10$ au at $R_0=100$ au \citep{Zhang21}. Here, we discuss how our results change when we vary the value of $H_0$. We explore a relatively wide range in values for $H_0/R_0$, from 0.08 up to 0.25, that we believe bracket sensible values. We keep the temperature power law index fixed at the literature { value $q=0.66$}. Fig. \ref{fig:honr_imlup} shows how the fitted parameters change with changing $(H/R)(R=100\mbox{au})$. Given that the sub-keplerian pressure term and the super-keplerian self-gravity term in the rotation curve provide opposite contributions, one naturally expects that the fitted disc mass increases with increasing $H/R$, which is indeed apparent from Fig. \ref{fig:honr_imlup}. It is however interesting to note that, unless $H/R$ becomes very large, the disc mass appears to not change significantly. For very large values of $H/R$, however, note that the fit to the rotation curve becomes progressively poorer. { The disk aspect ratio is linked to the disk temperature: in Fig. \ref{fig:honr_imlup}, the $H/R$ interval corresponds to a temperature range of [12K,140K] ad $R=100$au}.

\subsubsection{Using the quadratic method to retrieve the rotation curves}
\label{subsec:imlup_results}

We have also attempted to fit the data obtained by using the quadratic method for both \eddy and \discminer. In general, using the quadratic fit to the velocity results in a higher estimated disc mass, by roughly a factor 2. 
Indeed, using the SHO quadratic method with \eddy, we obtain $M_\star \approx 1.03 \, \text{M}_\odot$, $M_\text{d} \approx 0.19 \, \text{M}_\odot$ and $R_\text{c} \approx 87 \, \text{au}$, while for \discminer quadratic we obtain again very good agreement for the disc and stellar masses ($M_\star \approx 1.02 \, \text{M}_\odot$, $M_\text{d} \approx 0.23 \, \text{M}_\odot$), but we were not able to fit for the disc scale radius. 
{ A possible explanation for the discrepancy between Quadratic and Gaussian method could be attributed to the effect of the emission from the lower surface. In principle, since the quadratic method fits only the peak of emission, it should be less sensitive to the lower surface. However, it can happen that the emission coming from the lower surface is stronger compared to the upper one and in this case the quadratic fit leads to inaccurate results. As for the Gaussian method, the fits takes into account the whole emission line, not only the peak, and so the effect of the lower surface emission is to shift the fitted rotational velocity towards smaller values, underestimating it. A possible way to prevent this issue would be to fit with a double gaussian profile, in order to take into account both surfaces but we have not implemented it in this work.}

\subsubsection{Non axi-symmetric kinematics and spiral structure in IM Lup}

Recently, \citet{Verrios22} detected non-axisymmetric perturbations in the kinematics of IM Lup, attributing it to the presence of an embedded Jupiter mass planet. The appearance of velocity `kinks' \citep{Pinte18} or `wiggles' \citep{Hall20} is not unexpected in this source, given the known spiral morphology, that naturally has a kinematic counterpart (see also the analytical kinematical models of \citealt{Bollati21} and \citealt{Longarini21}, referring to a planetary spiral or a gravitational instability spiral, respectively). 

Firstly, we explicitly note that our model employs the azimuthally averaged rotation curve and we thus do not model in any way possible non-axisymmetric deviations from Keplerianity (although it would be interesting to ascertain how the appearance of non-axisymmetric perturbations affects the very determination of the rotation curve, a topic that we defer to further studies). 

In order to reproduce the observed scattered light images, \citet{Verrios22} require sub-micron dust grains to be well coupled to the gas, in turn implying a relatively high disc mass. Interestingly, their best model has a disc mass of $0.1M_{\sun}$, in perfect agreement with the one we derive here based on the rotation curve. \citet{Verrios22} note that the disc mass is high enough that self-gravity may have some dynamical effect in IM Lup, but they disregard the hypothesis that it is responsible for the spiral morphology and the associated velocity perturbations. However, our model has a $Q$-profile that is very close to unity across the disc extent and we thus expect the disc to be marginally stable, possibly giving rise to the observed spiral structure and observed kinematic wiggle.

\subsection{GM Aur}

\label{subsec:gmaur_results}

In Fig \ref{fig:rotcurves_gm_aur} we show the results of our fit to the GM Aur rotation curve obtained with \discminer. In this case, fitting a single model to both molecules appears to be more difficult, since the rotation traced by $^{12}$CO appears to be systematically lower than that of $^{13}$CO, a shift that cannot be simply accounted for by the difference in emitting height, which is included in our modeling. As a result, our best fit model lies in between the data of the two isotopologues, providing a relatively poor fit to each of the two. The disc mass is $M_{\rm d}\approx 0.26\, \mathrm{M}_{\odot}$, the stellar mass is $M_{\star}\approx 0.7\, \mathrm{M}_{\odot}$ and the scale radius is $R_{\rm c}\approx 60$ au. 

We also tried to fit the rotation curves traced by the two molecular lines individually, and the results are shown in Fig. \ref{fig:gmaur_12}. The resulting values for the star mass, disc mass and scale radius are: $M_{\star}\approx 0.79\, \mathrm{M}_{\odot}$, $M_{\rm d}\approx 0.11\, \mathrm{M}_{\odot}$ and $R_{\rm c}\approx 65$ au for the $^{12}$CO line fit and a rather implausible $M_{\star}\approx 0.6\, \mathrm{M}_{\odot}$, $M_{\rm d}\approx 0.44\, \mathrm{M}_{\odot}$ and $R_{\rm c}\approx 59$ au for the $^{12}$CO line. 

The origin of the mismatch between the rotation curve traced by the two molecules cannot simply be the effect of the different height of the emitting layers, as discussed above, but might arise from a vertical stratification in temperature. Indeed,  \citet{Law21} describe the 2D structure of the disc, noting that the temperature of the $^{12}$CO line is systematically higher than that of $^{13}$CO. However, this effect also cannot explain the observed difference in rotation velocity. Indeed, approximately, we can write the rotational velocity of the two isotopologues (including here for simplicity only the pressure gradient terms) as:
\begin{equation}
    v_{12}^2 \approx v_{\rm K}^2 - \gamma^\prime c^2_{\rm s,12},
\end{equation}
and
\begin{equation}
    v_{13}^2 \approx v_{\rm K}^2 - \gamma^\prime c^2_{\rm s,13},
\end{equation}
where $c_{\rm s,12}$ and $c_{\rm s,13}$ are the sound speeds corresponding to the temperatures of the two isotopologues. \citet{Law21} parameterize the gas temperatures of the two molecules as:
\begin{equation}
    T_{12} = 52\mbox{K}\,\left(\frac{R}{100\mbox{au}}\right)^{-0.61},
\end{equation}
and
\begin{equation}
    T_{13} = 22\mbox{K}\,\left(\frac{R}{100\mbox{au}}\right)^{-0.26}.
\end{equation}
The difference between the two rotation curves can thus be written as:
\begin{equation}
    \frac{v_{13}^2-v_{12}^2}{v_{13}^2}\approx \gamma'\left(\frac{c_{\rm s,13}}{v_{\rm K}}\right)^2\left(\frac{T_{12}}{T_{13}}-1\right).
    \label{eq:Tstrat}
\end{equation}
In Fig. \ref{fig:dv} we plot the difference between the two rotation curves obtained from \discminer (blue dots) and the expected difference due to the temperature stratification according to Eq. (\ref{eq:Tstrat}) (orange line). We thus see that the vertical stratification would produce a mismatch that is one order of magnitude smaller than observed and thus cannot explain it.

The observed difference can thus be due to two facts: either we do not recover correctly the rotation curve in this specific source, which may be due to an incorrect determination of the height of the emitting layer in the \citet{Law21}, or the outer disc is not in exact centrifugal balance. We know that the outer disc of GM Aur is highly perturbed \citep{Schwarz21} and the gas kinematics might thus not trace pure rotational motions. Thus, our results for the GM Aur disc should be regarded with some caution.

\section{Conclusions}

\label{sec:conclusions}

In this paper we have analysed the rotation curves of two protostellar discs, IM Lup and GM Aur, for which high quality kinematical data are available from the MAPS program. In particular, we have refined our technique to infer dynamically the disc mass, by fitting the observed rotation curve to a model curve that includes the disc self-gravity. We had previously used the same method to determine the mass of the disc in Elias 2-27 \citep{Veronesi21} and here we thus extend the same analysis to two additional sources. 

With respect to our results for Elias 2-27 \citep{Veronesi21}, we have improved the analysis in several respects:
\begin{enumerate}
    \item We retrieve the rotation curve by using the \discminer tool \citep{Izquierdo21}. 
    \item We have tested the robustness of the rotation curve retrieval by comparing the results of \discminer to those obtained from the \eddy package \citep{teague2019eddy}. 
    \item In modeling the rotation curve, we include the contribution of the pressure gradient, including its vertical dependence (but retaining the assumption of vertical isothermality).
    \item We also assume the disc surface density to be described by a tapered power-law and in this way we are also able to fit dynamically for the tapering radius $R_{\rm c}$, which we were not able to do in \citet{Veronesi21}.
\end{enumerate}
One additional refinement to the model (not included in the present work) is to also include vertical temperature stratification in the modeling of the rotation curve, that we defer to subsequent work.

Our results are the following:

{\bf IM Lup.} For this source, we obtain reliable rotation curves both with \discminer and \eddy and the two curves are consistent with each other. We fit the stellar mass to $M_\star \approx 1\, \mathrm{M}_{\odot}$, the disc mass to $M_{\rm d} \approx 0.1\, \mathrm{M}_{\odot}$ and the tapering radius to $R_{\rm c}\approx 70-90$ au. Our resulting disc mass implies a disc/star mass ratio of $\approx 0.1$, and we estimate the gravitational stability parameter $Q$ to become $\sim 1$ in the outer disc, which points to a gravitational instability as the origin of the spiral structure observed in the dust morphology \citep{Huang18}. The resulting gas/dust ratio turns out to be $\approx 60$. This fit is robust when comparing the rotation curves obtained with \discminer and with \eddy. We have also analysed the robustness of our results with respect to the chosen value of the pressure scale-height $H/R$, finding a relatively weak dependence, at least for small $H/R$. When $H/R$ becomes very large, however, the best fit disc mass increases, but the fit becomes progressively poorer. 

We have used the Gaussian method in \eddy and \discminer to obtain the velocity from the emission line data, which provides a smoother rotation curve. We have also tested the effect of using a quadratic fit to the line profile instead, which generally gives poorer results for the rotation curve. However, also in this case, \eddy and \discminer provide consistent results, although the fitted disc masses appear to be a factor $\sim 2$ larger than the ones obtained with the Gaussian profile.

{\bf GM Aur.} This source is more difficult to model in several respects. Firstly, \eddy and \discminer do not provide consistent results. Contamination from emission from the back side of the disc is particularly evident in this source and prevents \eddy from retrieving accurately the rotation curve. Additionally, the curves referring to the two isotopologues considered appear not to be consistent with each other, a difference that cannot be attributed by the difference in emitting scale height nor to a vertical temperature stratification. As a result, a single model fitting both rotation curves is more difficult to find. Our best fit model for both species has a stellar mass of $M_\star \approx 0.7\, \mathrm{M}_{\odot}$, a disc mass of $M_{\rm d} \approx 0.26\, \mathrm{M}_{\odot}$ and a tapering radius of $R_{\rm c}\approx 60$ au. However, given the difficulties mentioned above and the likely possibility that the molecular data do not trace perfectly the rotation curve, given the observed interaction with the infalling cloud \citep{Schwarz21}, we should consider our fit parameters with caution. 

Our work demonstrates that, with the currently available precision in gas kinematics in discs, it is possible to provide { reliable} dynamical measurements of the disc mass, at least for discs whose kinematics is not perturbed by the environment and that have a high enough mass to provide sizable deviations from Keplerianity. This measurement, however, is difficult, and essentially depends on the differential rotation between the inner and outer disc. As such, to this goal it is essential to have well sampled rotation curves, with high spatial resolution (to distinguish the inner and the outer disc) and good spectral resolution (to accurately measure the small deviations from Keplerianity in the outer disc). In addition, it is very important to have a good knowledge of the temperature structure of the disc (to correctly evaluate the impact of pressure support against gravity), in principle including vertical stratification, and of the height of the emitting layer. { Moreover, having a good knowledge of the disc structure is crucial. Indeed, the height of the emitting layer $z(R)$ affects the retrieval of the rotation curve. In addition, a possible disk warp could affect the projected velocities \citep{rosenfeld}: as far as the two analysed discs are concerned, there is no evidence of disc warping.} In the present work, we have simply assumed these parameters from the literature, but in the future one might want to construct a single framework to simultaneously fit the whole disc kinematical structure.

New ALMA surveys specifically aimed at disc kinematics will provide additional data onto which to perform the analysis outlined here and we thus expect the sample of discs with dynamically measured mass to increase in the upcoming years.

\section*{Acknowledgements}

This paper makes use of the following ALMA data: ADS/JAO.ALMA\#2018.1.01055.L. ALMA is a partnership of ESO (representing its member states), NSF (USA), and NINS (Japan), together with NRC (Canada), NSC and ASIAA (Taiwan), and KASI (Republic of Korea), in cooperation with the Republic of Chile. The Joint ALMA Observatory is operated by ESO, AUI/NRAO, and NAOJ. This work has received funding from the European Union’s Horizon 2020 research and innovation programme under the Marie Sklodowska-Curie grant agreement No 823823 (RISE DUSTBUSTERS project). We used the emcee algorithm (Foreman-Mackey et al. 2013), the corner package Foreman-Mackey (2016) to produce corner plots and python-based MATPLOTLIB package (Hunter 2007) for all the other figures. This work was partly supported by the Italian Ministero dell’Istruzione, Università e Ricerca through the grant Progetti Premiali 2012-iALMA (CUP C52I13000140001), by the Deutsche Forschungsgemeinschaft (DFG, German Research Foundation) - Ref no. 325594231 FOR 2634/2 TE 1024/2-1, by the DFG Cluster of Excellence Origins (www.origins-cluster.de). This project has received funding from the European Research Council (ERC) via the ERC Synergy Grant ECOGAL (grant 855130). BV acknowledges funding from the ERC CoG project PODCAST No 864965.

\section*{Data Availability}

The kinematical data of the CO molecules used in this paper are publiclly avaiable from the MAPS survey \url{https://alma-maps.info/data.html}. 



\bibliographystyle{mnras}
\bibliography{example} 




\appendix

\section{Pressure gradient contribution to the rotation curve}
\label{appendixA}

We start form hydrostatic equilibrium in the vertical direction:
\begin{equation}
\frac{1}{\rho}\frac{\mbox{d}P}{\mbox{d}z}=-\frac{GM_\star}{(R^2+z^2)^{3/2}}z.
\end{equation}
In the following, we will assume the disc to be vertically isothermal, but radially varying, so that
\begin{equation}
c_{\rm s}^2\propto T \propto R^{-q}.
\end{equation}
Hydrostatic equilibrium can be re-written as:
\begin{equation}
c_{\rm s}^2\frac{\mbox{d}\log\rho}{\mbox{d}z}=-\frac{GM_\star}{(R^2+z^2)^{3/2}}z,
\end{equation}
whose solution is:
\begin{equation}
\rho(R,z)=\rho_0(R)\exp\left[-\frac{R^2}{H^2}\left(1-\frac{1}{\sqrt{1+z^2/R^2}}\right)\right],
\label{eq:appendixdensity}
\end{equation}
where $H=c_{\rm s}/\Omega_{\rm K}$ is the pressure scale-height and $\Omega_{\rm K}=\sqrt{GM_\star/R^3}$ is the Keplerian angular velocity.
We also assume that $\rho_0$ is a power-law in $R$. Specifically, if $\Sigma\propto \rho_0H\propto R^{-\gamma}$, we have that  $\rho_0\propto R^{(q/2-3/2-\gamma)}$. With these definitions, we also have that the pressure at the midplane $P_0\propto R^{-\gamma'}$, where $\gamma'=3/2+\gamma+q/2$, and
\begin{equation}
P(R,z)=P_0(R)\exp\left[-\frac{R^2}{H^2}\left(1-\frac{1}{\sqrt{1+z^2/R^2}}\right)\right].
\end{equation}
Note that the vertical structure of the density and pressure correctly reduces to the usual Gaussian $\exp(-z^2/2H^2)$ for $z\ll R$. 

\begin{figure*}
  \centering
		\subfigure[1][Not aligned spectra.  ]{\label{fig:im13_120_a}\includegraphics[width=0.25\paperwidth]{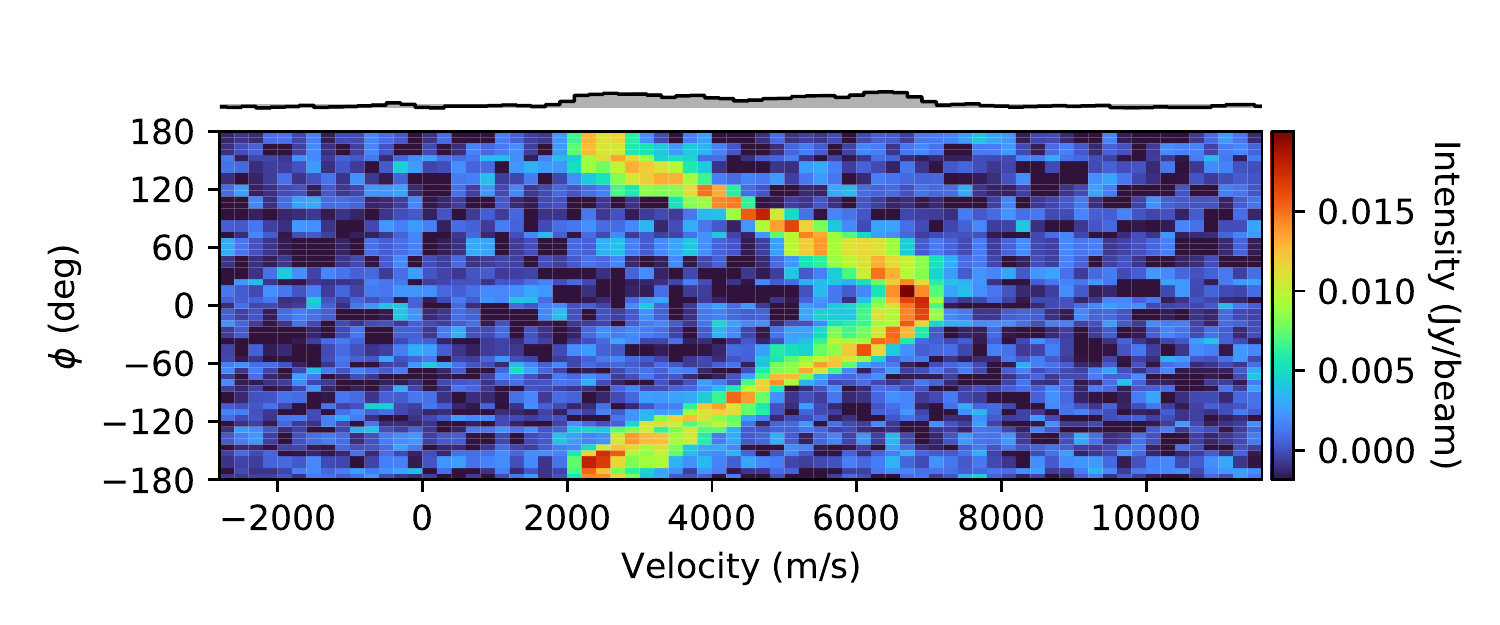}}
		\subfigure[2][Aligned spectra.]{\label{fig:im13_120_b}\includegraphics[width=0.25\paperwidth]{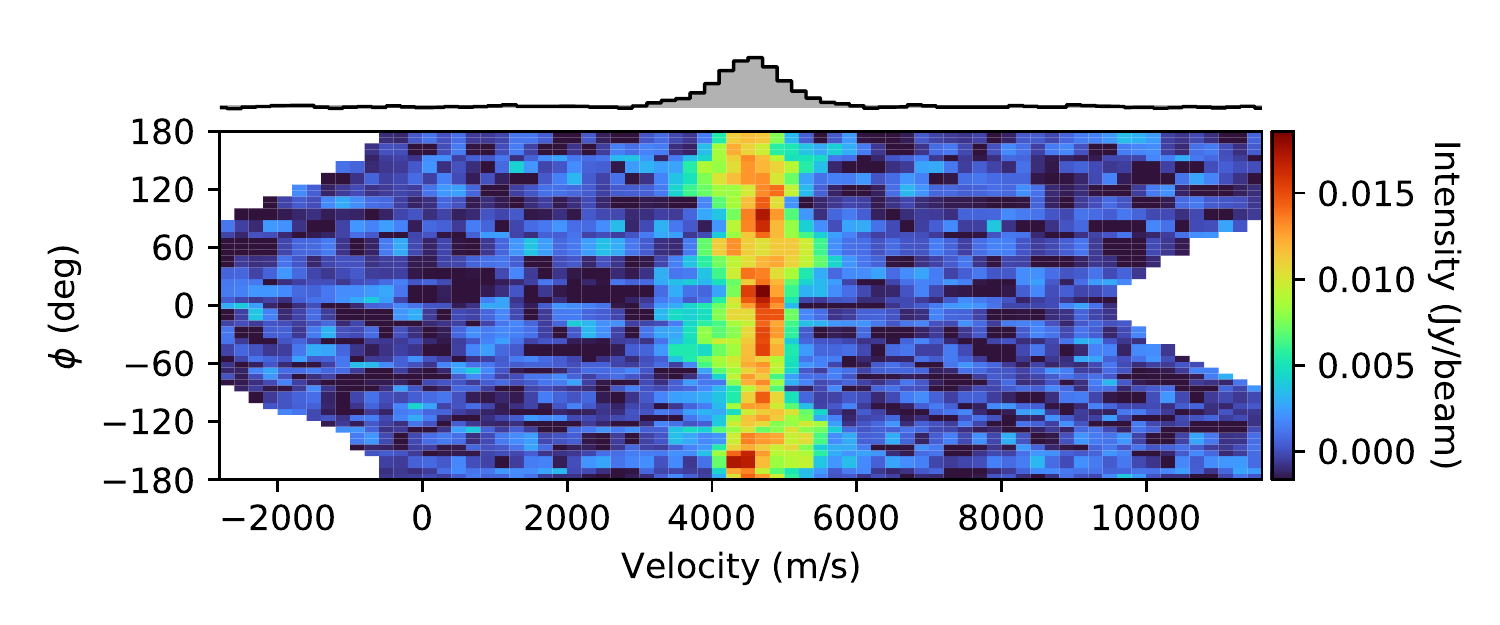}}
		\subfigure[3][Line centroids.]{\label{fig:im13_120_c}\includegraphics[width=0.25\paperwidth]{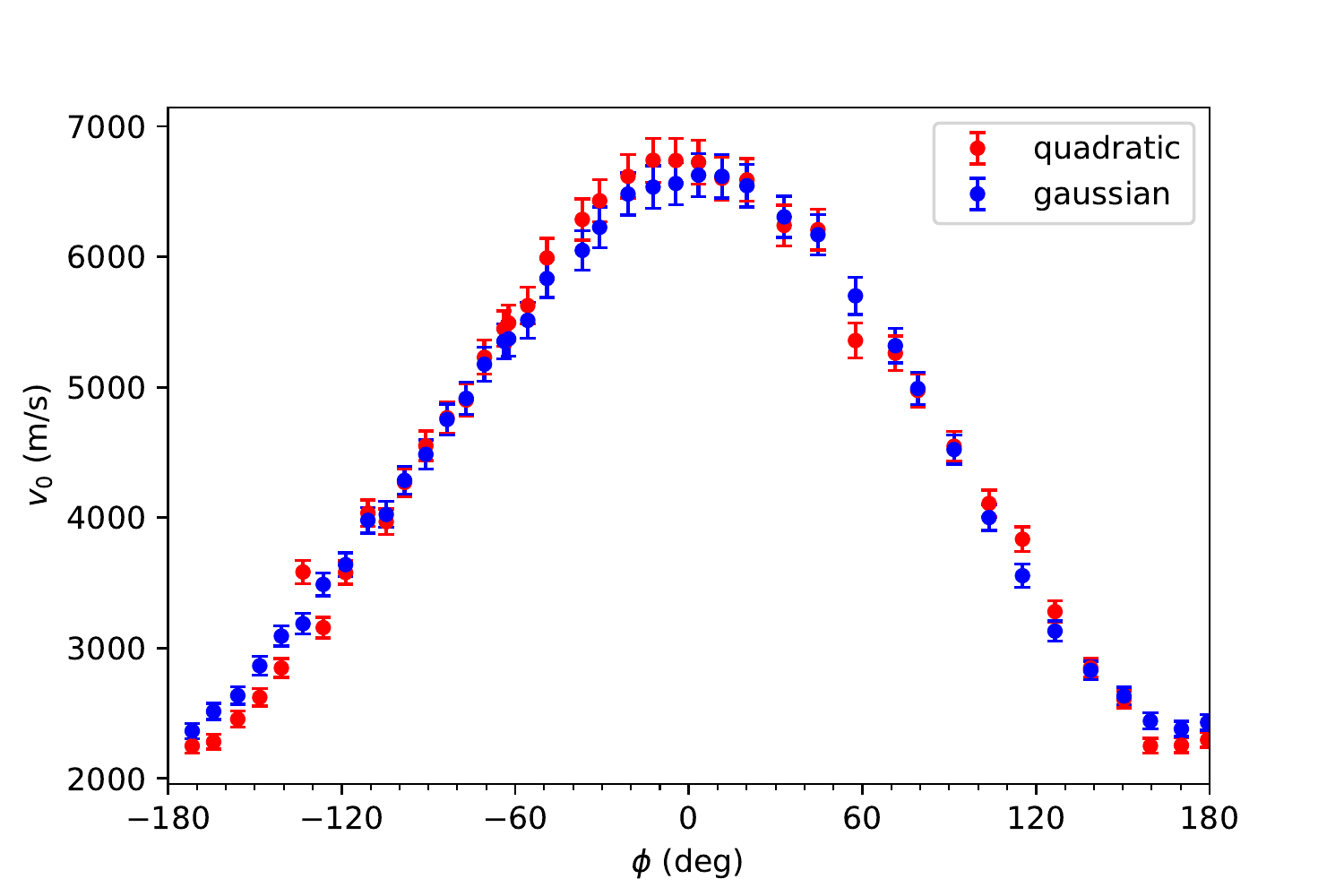}}
		
		\subfigure[4][Not aligned spectra.  ]{\label{fig:im120_a}\includegraphics[width=0.25\paperwidth]{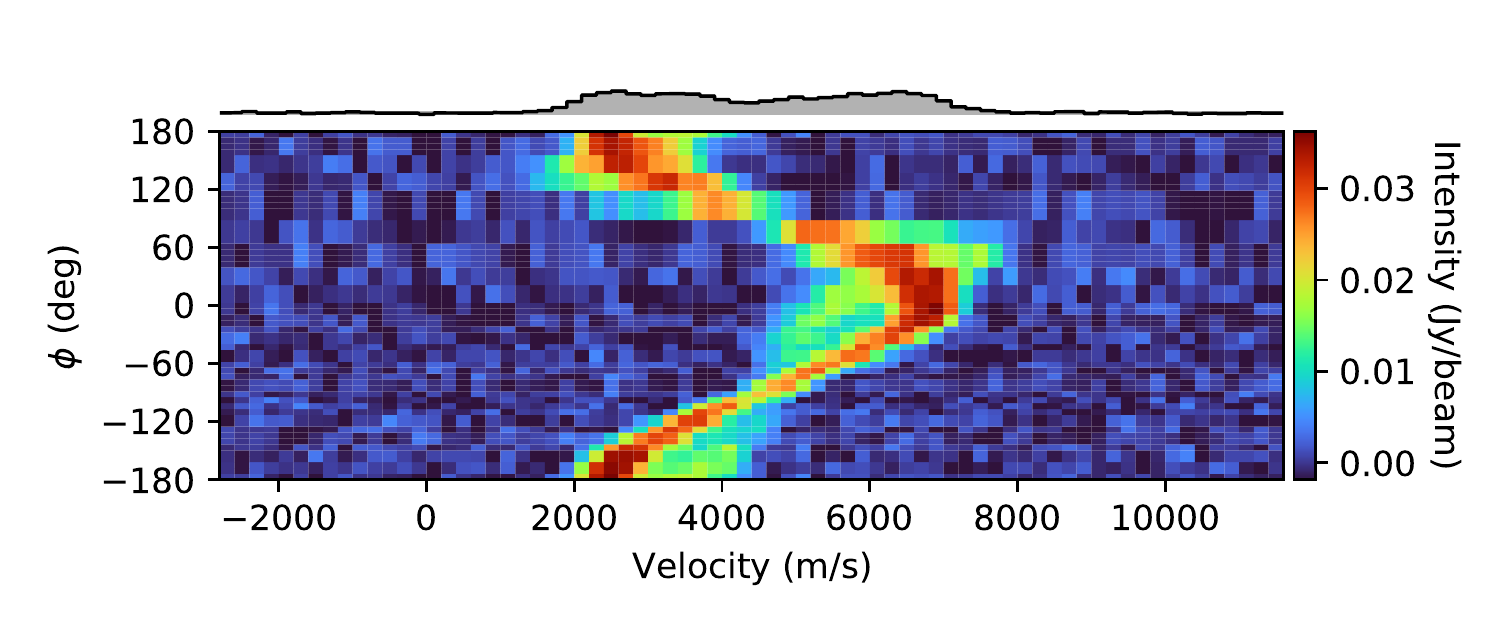}}
		\subfigure[5][Aligned spectra.]{\label{fig:im120_b}\includegraphics[width=0.25\paperwidth]{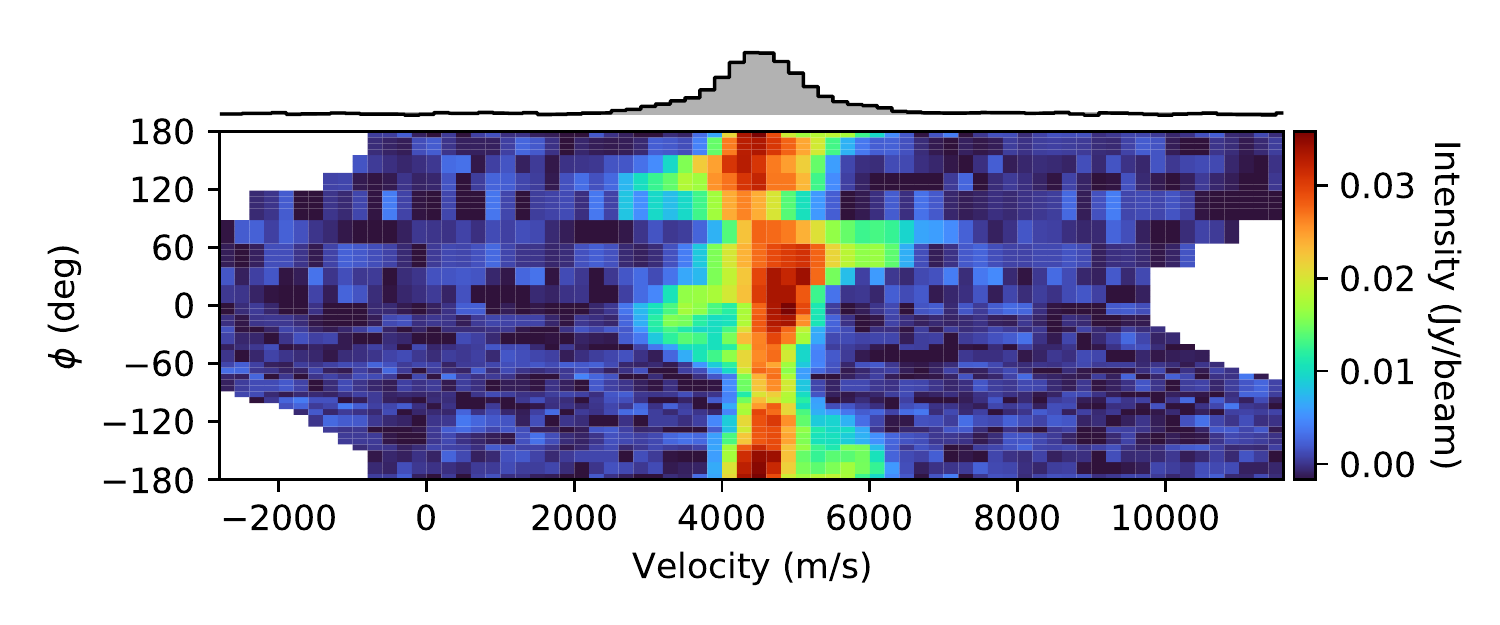}}
		\subfigure[6][ Line centroids ]{\label{fig:im120_c}\includegraphics[width=0.25\paperwidth]{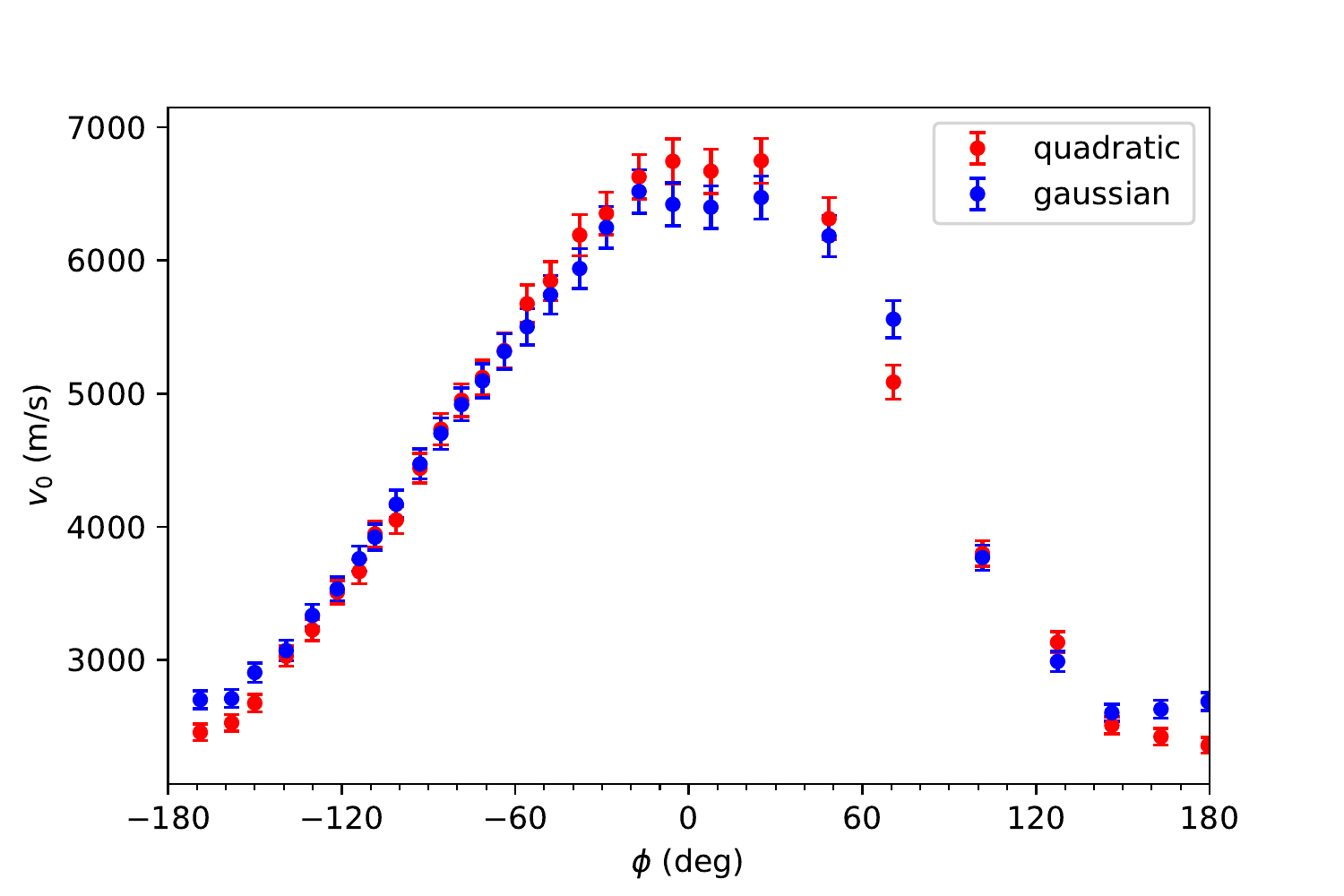}}
		
		\subfigure[7][Not aligned spectra. ]{\label{fig:gm120_a}\includegraphics[width=0.25\paperwidth]{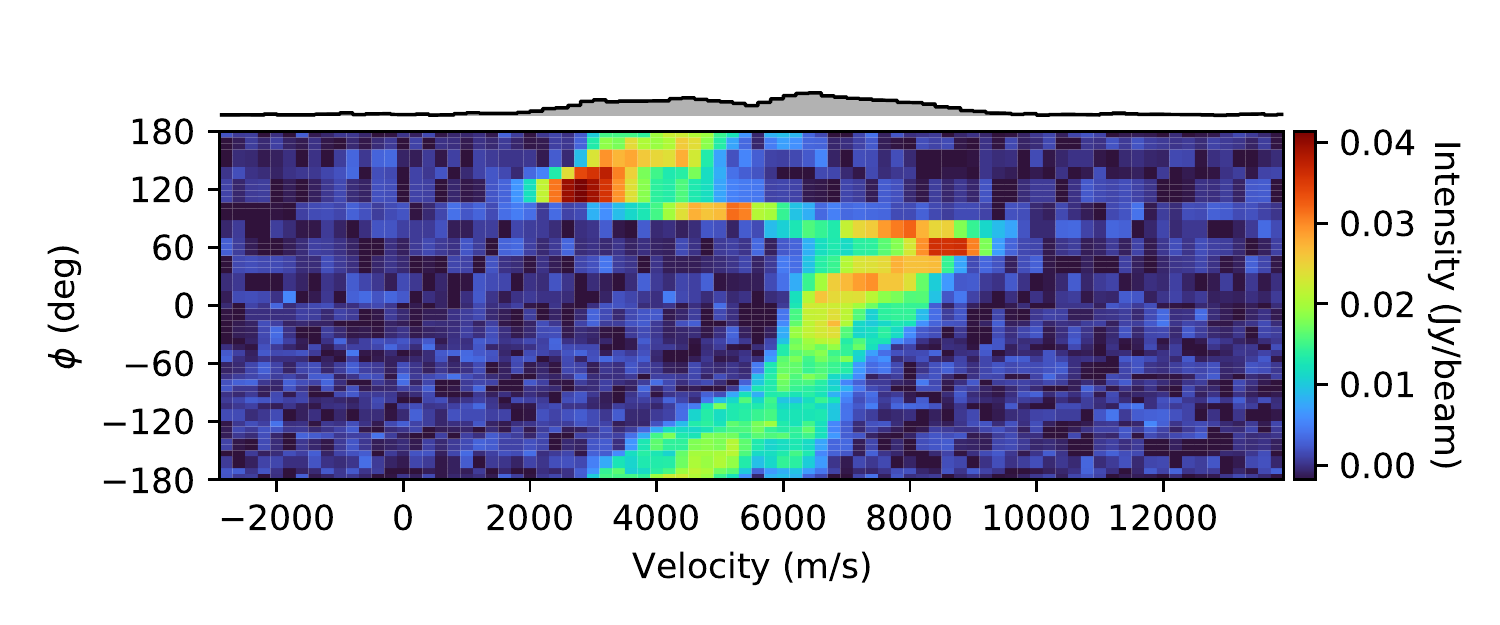}}
		\subfigure[8][Aligned spectra.]{\label{fig:gm120_b}\includegraphics[width=0.25\paperwidth]{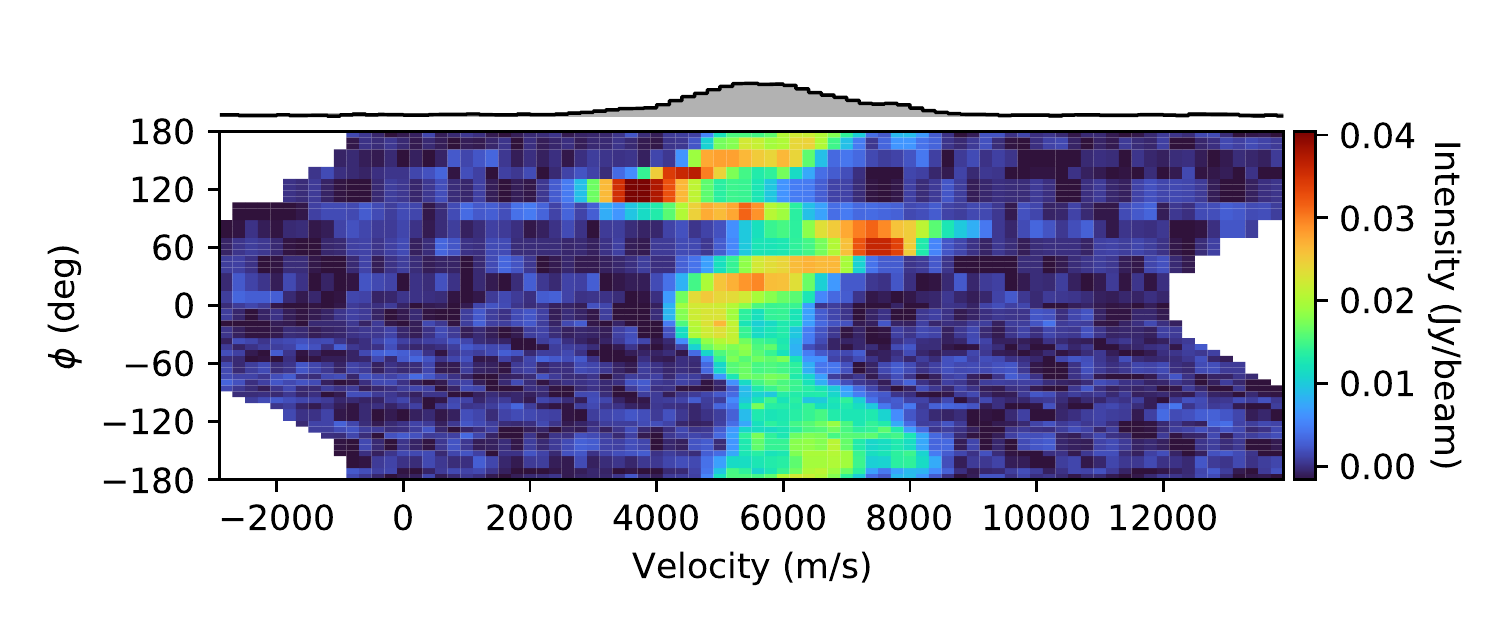}}
		\subfigure[9][Line centroids.]{\label{fig:gm120_c}\includegraphics[width=0.25\paperwidth]{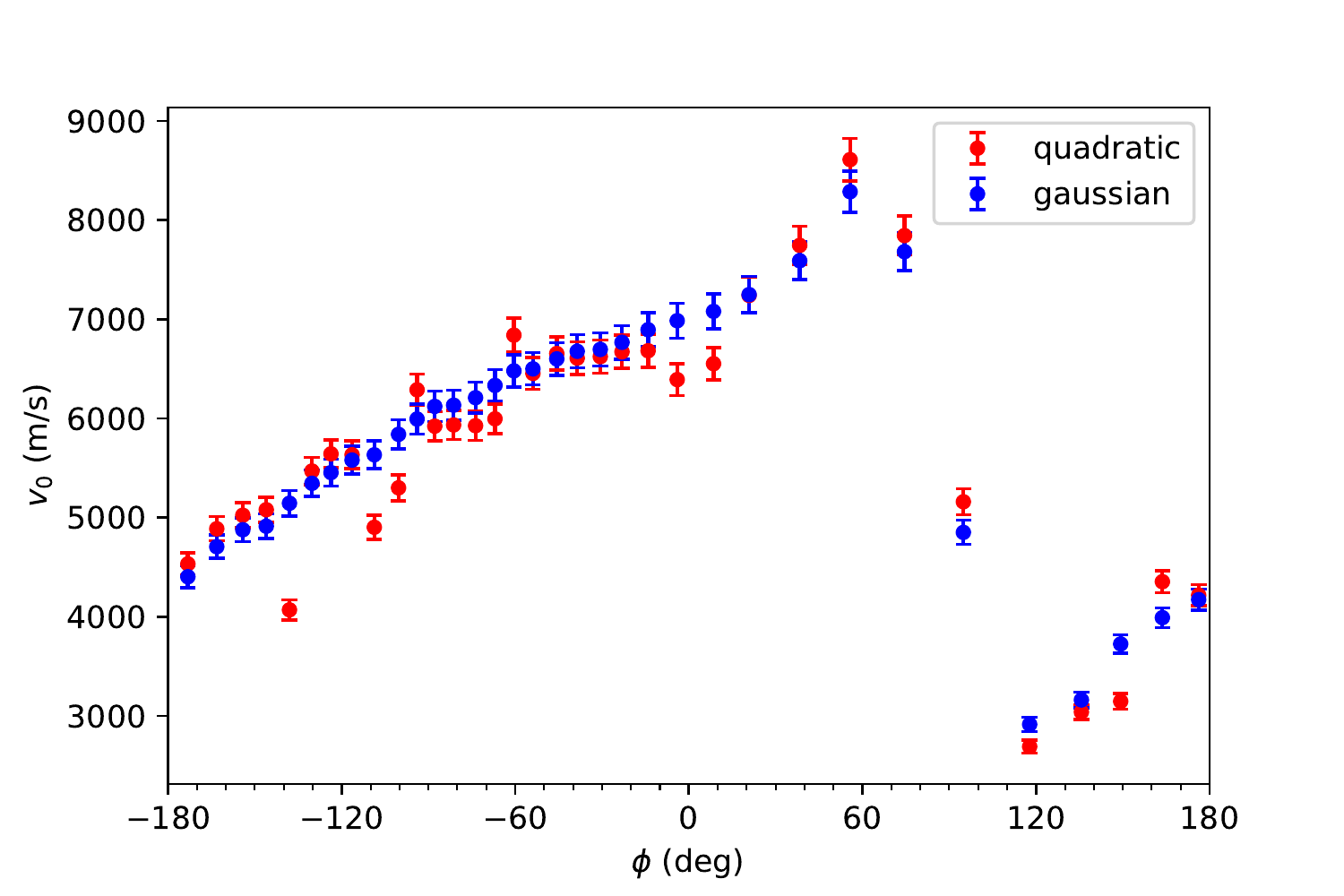}}
		
  \caption{First row: spectra of IM Lup \col for the annulus at 120 au. Second row: spectra of IM Lup \co for the annulus at 120 au.
  Third row: spectra of GM Aur \co for the annulus at 120 au.
  Left panels show the spectra at each azimuthal position, stacked on top of one another. The effect of rotation is highlighted by the shift in the peak intensity position for each spectrum along the velocity axis (the red shifted axis corresponds to $\phi = 0^{\circ}$). Middle panels show the straighten out spectra obtained by correcting the not aligned ones for the factor ${v}_{\mathrm{rot}}\cdot \cos (\phi )$, fitted with the quadratic method. On the top of the left and middle panels the azimuthally averaged spectrum is represented: when the correct rotation velocity is found, it reaches its maximum peak. Lastly, the right panels show the line centroids, i.e. the line of the peak intensity for each azimuthal spectrum, identified with the quadratic (red points) and the gaussian (blue points) method. 
  From the comparison of panels (a) and (d) we can see that the \col is less contaminated by the back side with respect to \co, as the emission surface is lower over the midplane. This is true both for the IM Lup and the GM Aur discs.
  From the comparison of panels (e) and (h) we can see that, while for IM Lup the front side dominates over the back side and the spectra are correctly aligned, for GM Aur the signals are comparable, resulting in misaligned spectra.}
  \label{fig:im12_50}
\end{figure*}

The pressure contribution to the rotation curve is
\begin{equation}
v_p^2=\frac{R}{\rho}\frac{\mbox{d}P}{\mbox{d}R}.
\end{equation}
This can be easily obtained from the structure of the disc outlined above, after appropriate differentiation. The result is:
\begin{eqnarray}
v_p^2  =  v_{\rm K}^2\left\{-\gamma'\left(\frac{H}{R}\right)^2+\frac{2}{(1+z^2/R^2)^{3/2}} \left[1+\frac{3z^2}{2R^2}-\left(1+\frac{z^2}{R^2}\right)^{3/2} \right. \right.\\
\nonumber \left.\left.  -\frac{\mbox{d}\log H}{\mbox{d}\log R}\left(1+\frac{z^2}{R^2}-\left(1+\frac{z^2}{R^2}\right)^{3/2}\right)\right]\right\},
\end{eqnarray}
where $v_{\rm K}=\Omega_{\rm K}R$.
Finally, including also the stellar gravitational field (but neglecting self-gravity), the rotation curve of the disc is  
\begin{eqnarray}
v_{\rm rot}^2  = v_{\rm K}^2  \frac{1}{(1+z^2/R^2)^{3/2}}+v_{\rm p}^2 = \\
\nonumber v_{\rm K}^2 \left\{-\gamma'\left(\frac{H}{R}\right)^2 +\frac{2}{(1+z^2/R^2)^{3/2}} \left[\frac{3}{2}\left(1+\frac{z^2}{R^2}\right) - \right. \right.\\
\nonumber \left.\left. -\left(1+\frac{z^2}{R^2}\right)^{3/2}  -\frac{\mbox{d}\log H}{\mbox{d}\log R}\left(1+\frac{z^2}{R^2}-\left(1+\frac{z^2}{R^2}\right)^{3/2}\right)\right]\right\},
\label{eq:appendixfull}
\end{eqnarray}
Under the above assumption on the radial profile of $T$,
\begin{equation}
\frac{\mbox{d}\log H}{\mbox{d}\log R}=3/2-q/2,
\end{equation}
and \citep{Nelson13}
\begin{equation}
    v_{\rm rot}^2  = v_{\rm K}^2 \left[1-\gamma'\left(\frac{H}{R}\right)^2 - q\left(1-\frac{1}{\sqrt{1+(z/R)^2}}\right) \right].
\end{equation}
Thus, for an isothermal disc (both radially and vertically), i.e. when $q=0$, equation (\ref{eq:appendixfull}) reduces identically to:
\begin{equation}
    v_{\rm rot}^2  = v_{\rm K}^2 \left[1-\gamma'\left(\frac{H}{R}\right)\right]
\end{equation}
and the disc rotates exactly on cylinders with no vertical dependence of $v_{\rm rot}$. The above result had been obtained previously only in approximate form by \citet{Fromang11} in the limit where $z\ll R$, but it holds exactly at any $z$ if one uses the correct expression for the vertical density structure for an isothermal disc (\ref{eq:appendixdensity}). 

One can expand the previous equations up to $O(z^4/R^4)$, obtaining approximate expressions for $v_{\rm p}$ and $v_{\rm rot}$:
\begin{equation}
v_p^2  =  v_{\rm k}^2\left\{-\gamma'\left(\frac{H}{R}\right)^2+\frac{\mbox{d}\log H}{\mbox{d}\log R}\left(\frac{z}{R}\right)^2 + O\left(\frac{z^4}{R^4}\right)\right\},
\end{equation}
(which is the result that one obtains when considering the simple Gaussian vertical profile for density and pressure) and
\begin{align}
v_{\rm rot}^2= v_{\rm K}^2 & \frac{1}{(1+z^2/R^2)^{3/2}}+v_p^2\approx\\
\nonumber v_{\rm k}^2  & \left[1-\gamma'\left(\frac{H}{R}\right)^2-\frac{q}{2}\left(\frac{z}{R}\right)^2 + O\left(\frac{z^4}{R^4}\right)  \right]
\end{align}

\section{Examples of rotation curve retrieval using Eddy}



\label{appendixB}

Here, we show three examples on how does \eddy perform in retrieving the rotation curve. We show examples from the IM Lup data set, at 120 au for the two molecules and for GM Aur, at 120 au for the \co line, see caption for details. As one can see, for IM Lup the fit with a cosine function to the peak emission is relatively good, but the aligned spectra show some evidence of the emission from the back side of the disc (seen as the green sinusoidal residual after alignment in the middle panels). Conversely, the case of GM Aur is more problematic, and in this particular annulus the shape of the azimuthal variation is not well represented by a cosine function.

\section{Numerical integration of the self-gravitating term}
{ The self-gravitating term of the rotation curve (Eq. \ref{sgterm}) can only be assessed numerically. To do so, we used \textsc{QUADPACK}, a Fortran library for numerical integration of one-dimensional functions \citep{quadpack}. In particular, we used a quadratic method of integration. Since the integration is computationally consuming, we integrated Eq. (\ref{sgterm}) on a two-dimensional grid (R,Z) and then, at each step of the fitting procedure, we interpolate the self-gravitating term from the grid, according to the value of $(R,z(R))$. }


\label{lastpage}
\end{document}